\documentclass[sigconf]{acmart} 
\renewcommand\footnotetextcopyrightpermission[1]{} 
\usepackage[utf8]{inputenc}
\makeatletter
\let\wfs@comment@comment\comment
\let\comment\@undefined
\usepackage{changes}
\let\wfs@changes@comment\comment
\let\comment\@undefined

\newcommand\comment{%
    \ifthenelse{\equal{\@currenvir}{comment}}
    {\wfs@comment@comment}
    {\wfs@changes@comment}%
}
\usepackage{tikz}
\usetikzlibrary{arrows,positioning,decorations.pathreplacing}
\usepackage{url}
\usetikzlibrary{shapes}
\usepackage{import}
\usepackage{xifthen}
\usepackage{pdfpages}
\usepackage{transparent}
\usepackage{xcolor}
\usepackage{balance}
\usepackage{xspace}
\usepackage{amsmath}
\usepackage{hyperref}
\usepackage{cleveref}
\usepackage{amsfonts}%
\usepackage{multirow}
\usepackage{graphicx}
\usepackage{booktabs}
\usepackage{caption}
\usepackage{enumitem}
\usepackage{adjustbox}
\usepackage{array}
\newcolumntype{R}[2]{%
    >{\adjustbox{angle=#1,lap=\width-(#2)}\bgroup}%
    c%
    <{\egroup}%
}

\usepackage{todonotes}

\setuptodonotes{fancyline, inline, color=blue!30}
\usepackage{dashbox}
\usepackage[n,landau,notions,ff,mm]{cryptocode}
\definecolor{gamechangecolor}{gray}{0.74}
\definechangesauthor[name=arthur, color=red]{arthur}

\definecolor{cblue}{rgb}{0.0, 0.28, 0.9}   
\definechangesauthor[name=Duc, color=cblue]{duc}

\newcommand*\bcircled[1]{\tikz[baseline=(char.base)]{
    \node[fill=black,text=white, shape=circle,draw=black,inner sep=.6pt] (char) {#1};}}

\usepackage[framemethod=TikZ]{mdframed}
\mdfsetup{
    font=\small,
    leftmargin=5pt,
    rightmargin=5pt,
    skipbelow=5pt,
    skipabove=5pt,
    innertopmargin=5pt,
    innerbottommargin=5pt,
    innerleftmargin=5pt,
    innerrightmargin=5pt,
    frametitlerule=true,
    frametitlealignment=\hspace*{0pt}
}
\title{AMR: Autonomous Coin Mixer with Privacy Preserving Reward Distribution}
\author{Duc V. Le}
\affiliation{%
 \institution{Purdue University}
 \country{}
}
\author{Arthur Gervais}
\affiliation{%
 \institution{Imperial College London}
 \country{}
}

\newcommand{\system}{\ensuremath{\mathsf{AMR}}\xspace}
\newcommand{\pparagraph}[1]{\smallskip \noindent \textbf{#1.}}
\newcommand{\bset}{\{0,1\}}

\newcommand{\FF}{\mathbb{F}}

\newcommand{\anoset}{\mathsf{AnomSet}^h}

\newcommand{\depset}{\mathsf{AnomSet}^h}

\newcommand{\zksnark}{zk-SNARK\xspace}

\newtheorem{definition}{Definition}
\newtheorem{claim}{Claim}

\newcommand{\idx}{\mathsf{index}}
\newcommand{\rt}{\ensuremath{\mathsf{root}_{dep}\xspace}}
\newcommand{\rtr}{\ensuremath{\mathsf{root}_{rwd}\xspace}}
\newcommand{\initialize}{\fun{Init}}
\newcommand{\mkverify}{\fun{Verify}}
\newcommand{\prove}{\fun{Prove}}
\newcommand{\update}{\fun{Update}}

\newcommand{\mpath}{\mathsf{path}}

\newcommand{\pcom}{\fun{P}_{\com}}   
\newcommand{\vcom}{\fun{V}_{\com}}   

\newcommand{\statement}{\textit{st}\xspace}

\newcommand{\zksetup}{\fun{Setup}}
\newcommand{\zkprove}{\fun{Prove}}
\newcommand{\zkverify}{\fun{Verify}}

\newcommand{\drootlist}{\mathsf{RootList}_{wdr,k}}
\newcommand{\dpl}{\ensuremath{\mathsf{DepositList}}\xspace}

\newcommand{\rwnl}{\ensuremath{\mathsf{NullifierList}}\xspace}
\newcommand{\wnl}{\ensuremath{\mathsf{NullifierList}}\xspace}

\newcommand{\currentrroot}{\ensuremath{\rtr^{\key{curr}}}\xspace}
\newcommand{\nextrroot}{\ensuremath{\rtr^{\key{next}}}\xspace}

\newcommand{\acceptdeposit}{\fun{AcceptDeposit}\xspace}
\newcommand{\issuewithdraw}{\fun{IssueWithdraw}\xspace}
\newcommand{\issuereward}{\fun{IssueReward}\xspace}

\newcommand{\witness}{\mathsf{wit}}
\newcommand{\cm}{\mathsf{cm}}

\newcommand{\ek}{\mathsf{ek}}
\newcommand{\vk}{\mathsf{vk}}

\newcommand{\sparam}{\ensuremath{1^\lambda}\xspace}

\newcommand{\fun}[1]{\textsc{#1}}

\newcommand{\key}[1]{\mathsf{#1}}
\newcommand{\sk}{\key{sk}\xspace}
\newcommand{\pk}{\key{pk}\xspace}

\newcommand{\CreateDTX}{\fun{CreateDepositTx}\xspace}
\newcommand{\CreateWTX}{\fun{CreateWithdrawTx}\xspace}
\newcommand{\CreateRTX}{\fun{CreateRedeemTx}\xspace}

\newcommand{\pp}{\key{pp}\xspace}

\newcommand{\sample}{\ensuremath{\xleftarrow{\$}}\xspace}

\newcommand{\xcoin}{\ensuremath{\mathsf{amt}}\xspace}

\newcommand{\amt}{\ensuremath{\gamma}\xspace}

\newcommand{\com}{\mathsf{com}\xspace}

\newcommand{\tx}{\texttt{tx}\xspace}
\newcommand{\txd}{\textsf{tx}_{dep}\xspace}
\newcommand{\txreward}{\textsf{tx}_{rwd}\xspace}
\newcommand{\txr}{\textsf{tx}_{rwd}\xspace}
\newcommand{\txw}{\textsf{tx}_{wdr}\xspace}

\newcommand{\set}[1]{\ensuremath{\{#1\}}\xspace}

\newcommand{\sn}{\ensuremath{\mathsf{sn}}\xspace}

\begin{document}
\begin{abstract}
  It is well known that users on open blockchains are tracked by an industry providing services to governments, law enforcement, secret services, and alike. While most blockchains do not protect their users' privacy and allow external observers to link transactions and addresses, a growing research interest attempts to design add-on privacy solutions to help users regain their privacy on non-private blockchains.

  In this work, we propose to our knowledge the first censorship resilient mixer, which can reward its users in a privacy-preserving manner for participating in the system. Increasing the anonymity set size, and diversity of users, is, as we believe, an important endeavor to raise a mixer's contributed privacy in practice. The paid-out rewards can take the form of governance tokens to decentralize the voting on system parameters, similar to how popular ``Decentralized Finance (Defi) farming'' protocols operate. Moreover, by leveraging existing Defi lending platforms, \system is the first mixer design that allows participating clients to earn financial interest on their deposited funds.

  Our system \system is autonomous as it does not rely on any external server or a third party. The evaluation of our \system implementation shows that the system supports today on Ethereum anonymity set sizes beyond thousands of users, and a capacity of over $66,000$ deposits per day, at constant system costs. We provide a formal specification of our \zksnark-based \system system, a privacy and security analysis, implementation, and evaluation with both the MiMC and Poseidon hash functions. 
\end{abstract}
\maketitle
\pagestyle{plain}

\section{Introduction}
\label{sec:introduction}

\sloppy{More than a decade after the emergence of permissionless blockchains,} such as Bitcoin, related work has thoroughly shown that the blockchain's pseudonymity is not offering its clients strong anonymity. Several works have therefore
attempted to both, deanonymize clients, cluster addresses~\cite{androulaki2013evaluating,gervais2014privacy} as well as to
build privacy solutions to protect the clients' privacy~\cite{miers2013zerocoin,sasson2014zerocash,zero-to-monero,ruffing2014coinshuffle,ruffing2017valueshuffle,heilman2017tumblebit}. Those existing privacy solutions can be categorized into two classes: 
\emph{(i)} a fundamental blockchain redesign to natively offer better privacy to clients,
and \emph{(ii)} add-on privacy solutions that aim to offer privacy for clients of existing, non-privacy-preserving blockchains.

This paper focuses on add-on privacy solutions that mix cryptocurrency coins within an anonymity set. One known problem of such mixers is that their provided privacy depends on the anonymity set size, i.e., on the protocol's number of active clients. Also, in those systems, to gain a certain degree of privacy, users often need to keep their digital assets locked in the system for a certain period before withdrawing. This locking period prevents users from performing any financial activities on those assets, i.e., there is an opportunity loss of investing the assets for a financial return.

Hence, this work's particular focus is to find new ways to incentivize clients to participate in the mixer. 
First, similar to popular ``DeFi farming'' protocols~\cite{yield-farming}, our system, called \system, chooses to reward mixer participants by granting governance tokens when a client deposits coins for at least time $t$ within the mixer. Naturally, the reward payout must remain privacy-preserving, i.e., a reward payment must be unlinkable to a deposit from the same client of the mixer. Clients can utilize the collected tokens to govern \system in a decentralized manner, without the need for an external server or centralized entity. Secondly, by leveraging existing popular lending platforms~\cite{aave,compound,yearn}, \system can allow clients to earn interest on their deposited funds. This approach makes \system the first mixer design that generates financial interest on participants' funds.
We hope that such a mixer attracts clients that are privacy-sensitive and interested in a token reward to maximize the anonymity set and client diversity within \system.

We formalize the \zksnark-based \system system, and implement the mixer in
$1,013$ lines of Solidity code. A deposit costs $1.2m$ gas ($31.95$ USD), while
a withdrawal costs $0.3m$ gas ($9.12$ USD), receiving a reward amounts to
$1.5m$ gas ($41.07$ USD)~\footnote{{Estimated using Ethererum price of
$\$380.4$ in 08/25/2020 14:39 UTC.}}~\footnote{Using the gas price of 70 Gwei.
1 GWei is $1\times 10^{-9}$ Ether.}. 
These numbers support a real-world deployment, that could support over $66,000$
deposits per day given Ethereum's transaction throughput (assuming no
withdrawals). The resulting anonymity set sizes, which can easily exceed
$1,000$ while operating at constant system costs, offer stronger privacy than,
e.g.\ the ring signature-based privacy solution~\cite{meiklejohn2018mobius},
whose costs scale linearly with the size of the anonymity set and are hence
practically capped at anonymity set sizes of $24$ {($8m$ gas for
withdrawing)}.

\pparagraph{Our contributions can be summarized as follows}
\begin{itemize}[leftmargin=*]
    \item We formalize and present a practical \zksnark based mixer
        \system, which breaks the linkability between deposited and withdrawn
        coins of a client on a smart contract enabled blockchain, and we provide a
        formal security and privacy analysis of the proposed system.
    \item To decentralize \system's governance and incentivise clients to join
        the system, we invent a privacy-preserving reward scheme for its
        clients. We believe that in practice, an incentive scheme would attract
        more and a wider variety of clients to such privacy solution, and hence
        contribute to a better anonymity for all involved clients.
    \item We leverage popular existing lending platforms~\cite{aave,compound} to
        propose the first autonomous decentralized on-chain mixer that allows users
        to earn interest on their deposited fund. This approach further incentivises
        users to keep their funds in the system.
    \item We implement \system and show that the system can be deployed and
        operated efficiently on a permissionless blockchain. A deposit into the
        system costs $1.2m$ ($31.95$ USD), a withdrawal costs $0.3m$ gas
        ($9.12$ USD) and collecting a reward costs $1.5m$ gas ($41.07$ USD) in
        transaction fees on the current Ethereum network.
        The anonymity set size of \system could grow to up to
        $2^{d}$~\footnote{$d$ is the depth of the Merkle tree}, while
        operating at constant system costs once deployed (we applied a Merkle
        depth tree of $d=30$ within this evaluation). Generating client-side
        zkSnark proofs costs $3.607$ seconds
        respectively on commodity hardware.
\end{itemize}

\pparagraph{Paper Organization}
\Cref{sec:preliminaries} outlines the necessary background before explaining an overview of \system in 
\Cref{sec:systemoverview}
\Cref{sec:genericconstruct} formally outlines the algorithms of \system, together with the desired security goals and threat model. 
\Cref{sec:detailedcontruction} provides a detailed description of \system. 
\Cref{sec:systemanalysis} discusses how \system achieves the security goals.
\Cref{sec:evaluation} presents an implementation and evaluation of \system. 
\Cref{sec:relatedwork} summarizes related work.
\Cref{sec:conclusion} concludes this paper.
\Cref{sec:discussion} outlines different applications of \system and discusses possible future works. 

\section{Preliminaries}
\label{sec:preliminaries}
In this section, we define several building blocks for \system.

\subsection{Background on Smart Contract Blockchains and lending platforms}
\label{sub:ethereum}
\pparagraph{Ethereum Blockchain} The Ethereum blockchain acts as a distributed virtual machine that supports
quasi Turing-complete programs. The capability of executing highly expressive
languages in those blockchains enables developers to create \emph{smart contract}.
The blockchain also keeps track of the state of every
account~\cite{wood2014ethereum}, namely \emph{externally-owned accounts (EoA)}
controlled by a private key, and \emph{contract account} own by contract's code.
Transactions from EoA determine the state transitions of the virtual machine. 
Transactions are either used to transfer Ether or to trigger the execution of
smart contract code. The costs of executing functions are expressed in terms of \emph{gas} unit.
In Ethereum, the transaction's sender is the party that pays for the cost of
executing all contract operations triggered by that transaction. 
For a more thorough background on blockchains, we refer the interested reader
to~\cite{bonneau2015sok, bano2019sok}.


\pparagraph{Lending platforms on Ethereum blockchain} 
Smart-contract-enabled blockchains like Ethereum give rise to many other decentralized financial (Defi) applications. Defi applications allow parties to participate in the financial market without relying on any trusted third party while retaining full custody of their funds. Defi applications appear in different forms, such as decentralized exchanges, lending platforms, or derivatives. At the time of writing, the Defi space accumulates over $10$bn dollars of digital assets, and hundreds of millions of dollars of assets are traded daily in those Defi platforms. 

For this work, we focus on existing lending protocols~\cite{aave,compound}. At its core, lending protocols let \emph{borrowers} acquire digital assets with a specified interest rate by placing upfront collaterals into the system. Later, to retrieve the collaterals, \emph{borrowers} need to pay back the borrowed funds along with an additional interest amount. Similarly, users also act as \emph{lenders} by depositing digital assets into the protocol, and the deposited amount will generate interest until users redeem those assets. Finally, the interest rates for borrowing and lending are determined by the state of the lending platforms. 
In this work, we are only interested in the depositing and redeeming functionalities of lending platforms.
\begin{definition}
    A lending protocol, $\Sigma$, reserves the following actions:
    \begin{itemize}[leftmargin=*]
        \item $\mathsf{amt}_{\Sigma}\leftarrow$ \fun{Deposit}$(\mathsf{amt})$ takes as input of $\mathsf{amt}$ of coins, and outputs a corresponding amount of $\mathsf{amt}_{\Sigma}$ tokens.
        $\mathsf{amt}_{\Sigma}$ tokens are minted upon deposits and sent to the depositor, and the value of $\mathsf{amt}_{\Sigma}$ increases over time. 
        \item $\mathsf{amt}+ R\leftarrow$ \fun{Redeem}$(\mathsf{amt}_{\Sigma})$ takes as input $\mathsf{amt}_{\Sigma}$ tokens, and deposits $\mathsf{amt}+R$ to the function invoker. 
        The interest amount $R$ is determined by protocol $\Sigma$.
    \end{itemize}
\end{definition}
This definition aims to capture a high-level overview of how the depositing and redeeming functionalities work in a lending platform. 
For a detailed constructions of each actions in these lending protocols, we refer interested readers to~\cite{aave,compound,yearn}.

\pparagraph{Governance Token and Yield Farming in Decentralized Finance (DeFi)}
Users of DeFi platforms are often awarded governance tokens for interacting or
providing liquidity to DeFi platforms. These tokens can for instance be used
for governance and value accrual/yield farming. Governance means that users can
use their tokens to vote for changes in the contract during its lifetime. In
term of value accrual, platforms~\cite{curve,compound} allow users to lock
their governance tokens in a pool to be eligible to obtain trading fees
collected by the DeFi platform. This approach allows a fair distribution of
protocol fees to users who take on the opportunity cost of holding the
governance tokens. In this work, we adapt a similar technique of having a
distribution pool to fairly distribute total accrued interest collected by the
mixer to users.  


\subsection{Cryptographic Primitives}
\pparagraph{Notation}
We denote by $\sparam$ the security parameter and by 
$\mathsf{negl}(\lambda)$ a negligible function in $\lambda$.
We express by $(\pk, \sk)$ a pair of public and secret keys.
Moreover, we require that $\pk$ can always be efficiently derived from $\sk$, 
and we denote $\fun{extractPK}(\sk)=\pk$ to be the deterministic function to derive $\pk$ from $\sk$. 
$k||r$ denotes concatenation of two binary string $k$ and $r$. 
We denote $\mathbb{Z}_{\geq a}$ to denote the set of integers that are
greater or equal $a$, $\set{a, a+1, \dots}$. We let PPT denote probabilistic polynomial time.
We use $\statement[a,b,c \dots]$ to denote an instance of the statement where $a,b,c\dots$ 
have fixed and public values. We use a shaded area \gamechange{$i,j,k$} to denote 
the private inputs in the statement $\statement[a,b,c; \gamechange{$i, j, k$}]$.
\pparagraph{Collision resistant hash function} 
a family $H$ of hash functions is collision resistant, 
iff for all PPT $\mathcal A$ given $h\sample H$, the probability that $\mathcal{A}$ finds $x, x'$, such that
$h(x)=h(x')$ is negligible.
we refer to the cryptographic hash function $h$ as a fixed function $h: \bset^* \to \bset^\lambda$. 
For the formal definitions of cryptographic hash function family, we refer reader to~\cite{Rogaway2004}.



\pparagraph{\zksnark}
A zero-knowledge Succint Non-interactive ARgument of Knowledge
(\zksnark) can be considered as ``succinct'' NIZK
for arithmetic circuit satisfiability. 
For a field $\FF$, an arithmetic circuit $C$ takes
as inputs elements in $\FF$ and outputs elements in $\FF$. 
We use the similar definition from Sasson \emph{et al.}'s Zerocash paper~\cite{sasson2014zerocash}
to define arithmetic circuit satisfiability problem.
An arithmetic circuit satisfiability problem of a circuit $C:\FF^n\times\FF^h\rightarrow \FF^l$
is captured by relation $R_C=\set{(\statement,\witness)\in \FF^n\times \FF^h: C(st,\witness) = 0^l}$; 
the language is $\mathcal L_C = \set{\statement \in \FF^n~|~\exists~\witness \in \FF^l~s.t~C(st,\witness) = 0^l}$. 
\begin{definition}
\zksnark for arithmetic circuit satisfiability is triple of efficient algorithms $(\zksetup,\zkprove,\zkverify)$: 
\begin{itemize}[leftmargin=*]
    \item $(\ek, \vk) \leftarrow \fun{Setup}(\sparam, C)$ 
    takes as input the security parameter and the arithmetic circuit $C$, outputs
    a common reference string that contains 
    the evaluation key $\ek$ later used
    by prover to generate proof, and the verification key $\vk$ later used by
    the verifier to verify the proof. The public parameters, $\pp$, is given implicitly
    to both proving and verifying algorithms.
    \item $\pi\leftarrow \fun{Prove}(\ek,\statement,\witness)$ takes as input the
    evaluation key $\ek$ and $(\statement, \witness) \in R_C$, outputs a proof 
    $\pi$ that $(\statement, \witness)\in R_C$
    \item $0/1\leftarrow\fun{Verify}(\vk, \pi, \statement)$ takes as input the verification
    key, the proof $\pi$, the statement $\statement$, outputs $1$ if $\pi$ is valid proof
    for $\statement \in \mathcal{L}_C$. 
\end{itemize}
\end{definition}
In additional to \emph{Correctness, Soundness,} and \emph{Zero-knowledge} properties, 
a \zksnark requires two additional properties \textit{Succinctness} and \textit{Simulation extractability}.
We defer the definitions of these properties to~\cite{groth-zksnark-2016}.



\pparagraph{Commitment Scheme}
A commitment scheme allows a client to commit to chosen values while keeping
those values hidden from others during the committing round, and later during
the revealing round, client can decide to reveal the committed value.  

\begin{definition}
A commitment scheme $\com=(\fun{P}_{\com}, \fun{V}_\com)$ consists of:
A \emph{committing} algorithm $\fun{P}_{\com}(m, r)$ takes as input a message $m$ and randomness $r$,
and outputs the commitment value ${c}$. 
A \emph{Reveal} algorithm, $\fun{V}_{\com}(c,m,r)$  takes as input a message $m$, 
and the decommitment value $r$ and a commitment $c$, and returns 1 iff $c=\fun{P}_{\com}(m,r)$. Otherwise, returns 0.
\end{definition}

We use commitment schemes that achieve two properties: \emph{binding}
means that given commitment $c$, it is difficult to find a different
pair of message and randomness whose commitment is $c$, and 
\emph{hiding} means that given commitment $c$, it is hard to learn anything 
about the committed message $m$ from $c$.

\pparagraph{Authenticated Data Structure (ADS)} 
An authenticated data structure can be used to compute a short digest of a set 
$X=\set{x_1, \dots, x_n}$, so that later one can prove certain properties of $X$
with respect to the digest. In this work, we are only interested in a data
structure for set membership:
\begin{definition}
An authenticated data structure for set membership 
$\Pi=(\initialize,$ $\fun{Prove},$ $\fun{Verify},$ $\fun{Update})$
is a tuple of four efficient algorithms:
\begin{itemize}[leftmargin=*]
    \item $y\leftarrow\initialize(\sparam, X)$ the initialization algorithm 
    takes as input the security parameter and the set $X=\set{x_1, \dots, x_n}$ 
    where $x_i \in \bset^*$, output $y\in \bset^\lambda$.
    \item $\pi \leftarrow \prove(i, x, X)$ takes as input an element $x\in
        \bset^*, 1\leq i \leq n$, and set $X$, outputs a proof that $x=x_i \in
        X$.
    \item $0/1\leftarrow \mkverify(i, x, y, \pi)$ takes as input $1 \leq i \leq
        n, x\in \bset^*, y\in \bset^\lambda$, and proof $\pi$, output $1$ iff
        $x=x_i\in X$ and $y=\initialize(\sparam, X)$. Otherwise, return 0.
    \item $y' \leftarrow \update(i,x,X)$ takes as input $1\leq i\leq n, x\in
        \bset^*$ and set $X$, output $y' = \initialize(\sparam, X')$ where $X'$
        is obtained by replace $x_i \in X$ with $x$.
\end{itemize}
\end{definition}
We require the ADS to be \emph{correct} and
\emph{secure}. We defer the formal definitions of these properties to 
Boneh and Shoup's book~\cite{boneh-shoup-book-2020}. Typical examples of
authenticated data structures are Merkle tree~\cite{Merkle1988} or
RSA Accumulators~\cite{rsa-acc-li-2007,rsa-acc-benaloh-2007}. 

\begin{figure}[t]
    \centering
    \includegraphics[width=\columnwidth]{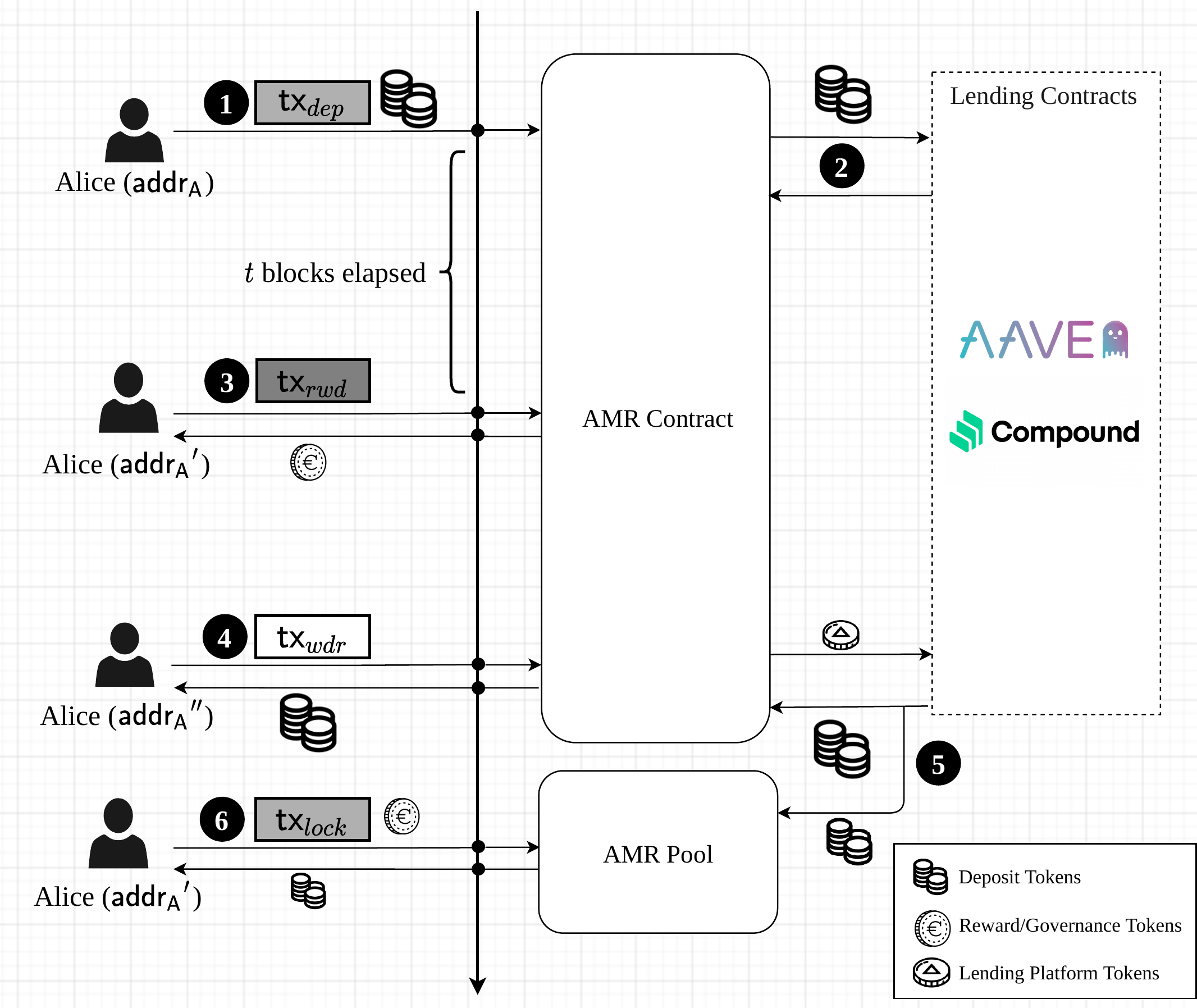}
    \caption{System Overview. In step~\protect\bcircled{1}, clients deposit coins
    to the \system contract.
    Upon receiving a valid deposit, \system deposit user's fund into lending platform~\protect\bcircled{2}. 
    In~\protect\bcircled{3}, a client proves to the contract to own a deposit older than $t$ blocks to obtain a reward.
    In~\protect\bcircled{4}, a client with a previous deposit 
    can withdraw the coin from the \system contract using a different address. 
    Upon receiving a valid withdrawing transaction, \system redeems user's deposit from 
    lending platform along with accrued interest in step~\protect\bcircled{5}, and deposits into
    user's address ($\mathsf{addr_A''}$). 
    In step~\protect\bcircled{6}, users with \system governance tokens can
    lock their tokens in an \system pool at any given time to obtain their
    share of the total accrued interest.
    }
    \label{fig:overview}
\end{figure}

\section{System Overview}
\label{sec:systemoverview}
We proceed to define the system components, overview, goals and the threat model.
\subsection{System Components}
There are three components of this system: the client, 
the \system smart contract, and onchain lending platforms.
    A \textbf{Client} interacts with the \system smart contract through
      externally owned accounts. A client can either deposit coins, withdraw
      coins, or redeem a reward.  
    The \textbf{\system Contract} is the blockchain smart contract that
    holds deposits, and handles withdrawals and reward redemptions. The
    contract keeps track of different data structures and parameters to
    verify the correctness and the integrity of transactions sent to the
    contract.
    The \textbf{\system Pool} is a smart contract that takes the accrued
    interest from a lending platform and proportionally distributes the reward among clients who lock their governance tokens to the pool.
    The \textbf{Lending Platforms} (cf.~\Cref{sub:ethereum}) are smart
    contracts that allow users to deposit digital assets and earn interest
    based on those assets.

\subsection{System Overview}

\Cref{fig:overview} outlines the overview of interactions in \system. 


\pparagraph{Deposits} 
In \system, clients deposit a fixed amount of coins into the system.
The client forms a depositing transaction to deposit coins, then sends this transaction through the P2P Network~\bcircled{1}. Once the transaction is validated, miners record the transaction in a blockchain block. Each deposit transaction decreases the balance of the clients' address by a fixed amount of coins. In step~\bcircled{2}, upon receiving a valid deposit from the user, \system deposits users fund
into lending platforms to obtain an equivalent amount of tokens for future withdrawals.

\pparagraph{Reward Redemptions} \system allows clients to earn governance
tokens as rewards based on certain conditions. In~\Cref{fig:overview}, the
requirement for a client to redeem a reward is to keep the deposit inside the
contract pool for $t$ blocks.  
To obtain a reward, a client forms a redeeming transaction and forwards the
transaction to the P2P Network~\bcircled{3}. 
The redeeming transaction includes
a cryptographic proof certifying that the client has deposited coins at least
$t$ blocks in the past and that the coins remain in the \system contract.
Finally, miners validate the redeeming transaction using the current state of
the \system contract. Once the redeeming transaction gets validated, the
transaction gets recorded to a blockchain block, and the network updates the
state of the \system contract.

\pparagraph{Withdrawals} 
The client forms a withdrawing transaction to withdraw coins, then sends this
transaction through the P2P Network~\bcircled{4}.
The withdrawing transaction includes cryptographic proof certifying that the
client has issued a depositing transaction in the past without revealing
precisely which one the depositing transaction is. 
In step~\bcircled{5}, upon receiving valid withdrawing transactions from the
client, the contract autonomously redeems the original deposit from lending
platforms and the accrued interest. 
Finally, the contract deposits the redeemed amount into user's address and the accrued interest into a separate \system pool.

\pparagraph{Fair Interest Allocation} 
At any given time, clients can lock their governance tokens to the \system pool~\bcircled{6}. \system distributes the
  total accrued interest to addresses that lock their \system governance tokens
  in \system pool. This step is straightforward, but offers a fair allocation of interest to users who contribute more to \system's privacy set.

\section{\system System}
\label{sec:genericconstruct}
In the following, we discuss various components of the \system system and provide more details of how \system operates.
In the following algorithm descriptions, we use $\texttt{tx}.\key{sender}$ to denote the address of the sender from which \texttt{tx} was sent.

\pparagraph{Condition for reward in \system} 
{
    %
    The longer time the clients wait before withdrawing/redeeming, 
    the more deposit transactions are issued to the \system contract.
    Thus, as the number of deposit transaction (i.e.\ the anonymity set) increases, the harder it is to link
    a withdrawing/redeeming transaction with the original deposit transaction.
    In \system, we incentivise clients by providing rewards to clients who can prove 
    that the deposit funds are not withdrawn before a certain time, measured in a number of blocks. 
    The provided reward can for instance represent a governance token for a
    client to participate in the decentralized governance of \system
    parameters.
}

\subsection{\system Contract Setup}
\label{sub:csetup}
The setup phase generates public parameters and data structures for the \system contract and clients. In particular, all cryptographic parameters are generated for the contract. The contract is also initialized with different data structures to prevent clients from double-withdrawal and double-redemption. The deposit and reward amounts, $\xcoin$ and $\xcoin_{rwd}$, are specified as a fixed deposit amount of coins and a fixed reward amount of governance tokens. 
The condition for redeeming rewards, $t_{con}$, is also declared. 
A lending platform, $\Sigma$, (cf.~\Cref{sub:ethereum}) is determined during this setup phase. 
A pool, $\Gamma_{\system}$, is deployed, and this pool periodically distributes the accrued interest to addresses that lock governance tokens.

We denote $\mathsf{pp}^h$ to be the state of the contract at block $h$. 
The state contains all data structures initialized during the setup phase. Moreover, this state is given implicitly to all clients' and contract's algorithms. Finally, the \system contract is deployed during this phase. 

\subsection{\system Client Algorithms}
\label{sub:clientalgorithms}
In our system, clients have access to the following algorithms to interact with the
\system smart contract. Also, all transactions are implicitly signed by the client using the
private key of the Ethereum account that creates the transaction.
\begin{itemize}[leftmargin=*]
	\item $(\witness,$ $\txd)$ $\leftarrow$ $\CreateDTX(\sk, \xcoin)$
    takes as input the private key $\sk$ and 
    the amount, $\xcoin$, coins specified in the setup phase, 
    outputs a deposit transaction $\txd$ and 
    the secret note $\witness$ which is used as witness for creating future
    withdraw and reward transactions. 
  \item $({\witness'}, \txreward) \leftarrow \CreateRTX(\sk', \witness)$ takes as
    input a private key $\sk'$ and the secret note $\witness$, outputs a
    reward-redeeming transaction $\txreward$ along with a new secret note, {$\witness'$}. 
  \item $\txw \leftarrow \CreateWTX(\sk'', \witness)$
    takes as input a private key $\sk''$ and the secret note $\witness$,
    outputs a withdrawing transaction $\txw$.
  \item $\tx_{lock}\leftarrow \fun{CreateLockTransaction}(\sk,\amt_{rwd}, t_{lock})$ takes
    as input an amount, $\amt_{rwd}$, of governance tokens and an unlock value, $t_{lock}$, 
    specifying how long, $\amt_{rwd}$, will remain locked in the system, outputs 
    a locking transaction, $\tx_{lock}$.
\end{itemize}

\subsection{\system Contract Algorithms}
\label{sub:contractalgorithms}

The \system contract should accept the deposit of funds, handle withdrawals, and reward redemptions.
Summarizing, the \system contract should provide the following functionalities.
\begin{itemize}[leftmargin=*]
    \item $0/1\leftarrow\acceptdeposit(\txd)$ takes as input the deposit
        transaction $\txd$. The \system contract deposits \xcoin into the lending platform 
        $\Sigma$ to obtain $\xcoin_{\Sigma}$.
        Finally, the algorithm outputs $1$ to denote a successful deposit, otherwise $0$. 

    \item $0/1\leftarrow\issuewithdraw(\txw)$ takes as input the withdraw
        transaction $\txw$. The \system contract uses $\xcoin_{\Sigma}$ to redeem $\mathsf{amt}+R$ from $\Sigma$.
        The algorithm outputs $1$ to denote a successful withdraw and
        deposits $\xcoin+R$ into $\txw.\key{sender}$. Otherwise, outputs 0.

    \item $0/1\leftarrow \issuereward(\txreward)$ takes as
        input the reward transaction $\txreward$ and the condition
        $t_\mathsf{con}$ specified during the setup algorithm, outputs $1$ if $\txreward$ satisfies the
        $t_\mathsf{con}$ for reward and deposit $\xcoin_{rwd}$ governance tokens as reward to
        $\txreward.\key{sender}$. Otherwise, output 0. 
\end{itemize}

\subsection{System Goals}
\label{sub:systemgoals}
\noindent In the following, we outline our system goals. 

\pparagraph{Correctness} Generally, \system needs to ensure that clients should not be able to steal coins from the \system contract or from other clients.  Moreover, we design \system such that clients can redeem a reward after they have deposited their coins into the \system contract for certain period of time, as a reward system will incentivise clients to deposit more into the system while contributing to the size of the anonymity set.

\system needs to provide the following guarantees: \emph{(i)} It is infeasible for clients to issue $n$ withdrawal transactions without issuing at least $n$ deposit transactions into the \system contract beforehand.  \emph{(ii)} It is infeasible for a client to issue a redeeming transaction without having any coins locked in the \system contract.  \emph{(iii)} A valid redeeming transaction indicates that a client always has at least one deposit locked in the \system contract for a specified duration.

\pparagraph{Privacy} In addition to correctness, \system needs to ensure the privacy to clients of the system. In particular, considering an adversary that has access to the history of all depositing, withdrawing, and redeeming transactions sent to \system contract, the system needs to ensure \emph{(i)} the unlinkability between deposit and withdrawing transactions \emph{(ii)} the unlinkability between deposit and redeeming transactions \emph{(iii)} the unlinkability between withdrawing and redeeming transactions.

\pparagraph{Availability} Similar to the availability definition proposed by Meiklejohn and Mercer's M\"{o}bius system~\cite{meiklejohn2018mobius}, \system should ensure that \emph{(i)} no one can prevent clients from using the mixer, and \emph{(ii)} once the coins are deposited to the contract, no one can prevent clients from withdrawing their coins.

\pparagraph{Frontrunning Resilience} 
Some transactions (i.e.\ deposit transactions) in \system alter the state of
the \system contract, while other transactions (i.e.\ withdrawing/redeeming
transactions) have to rely on the state of the contract to form the
cryptographic proofs.  Thus, if there are multiple concurrent deposit
transactions issuing to the contract, some transactions will get invalidated by
those transactions that modify the state of the contract.  
For example, in
\system, to withdraw or redeem a reward, a client Alice has to issue a
withdrawal and a redemption transactions that contains cryptographic proofs
proving that Alice deposited a coin in the past.  Alice generates those
cryptographic proofs w.r.t all current deposit transactions issued to the
\system contract. However, if another client Bob tries to deposit coins into
the \system contract, and Bob's transaction gets mined before Alice
withdrawing/redeeming transactions, the proofs included in Alice transactions
are no longer valid (because the state used for her proofs is outdated).
This is a \emph{front-running}
problem~\cite{zether-bunz-2020,eskandari-sok-frontrun-fc2019}.

Therefore, to ensure the usability of the system, the \system contract should
be resilient against \emph{front-running} by both clients and miners. 

\subsection{Threat Models}
\label{sub:threatmodels}
We assume that the cryptographic primitives (cf.\ \Cref{sec:preliminaries}) 
are secure. 
We further assume that adversaries are computationally bounded and can only
corrupt at most 1/3 of the consensus participants of the blockchain. 
Thus, we assume that an adversary cannot tamper with the execution of the
\system smart contract.  We assume that clients can always read the blockchain
state and write to the blockchain. Note that blockchain congestion might
temporarily affect the \emph{availability} property of \system, but does not
impact the \emph{correctness} and \emph{privacy} properties.
We assume that the adversary has the capabilities of a miner, i.e.\ can
reorder transactions within a blockchain block, inject its own transactions
before and after certain transactions.
Also, we assume that the adversary can always read all transactions issued to
the \system contract, while the transactions are propagating on the P2P
network, and afterwards when they are written to the blockchain. For a
withdrawal and a redeem transaction, we assume that the client pays transaction
fees either through a non-adversarial relayer (cf.\
Section~\ref{sec:discussion}), or the client possesses a blockchain address
with funds that are not linkable to his deposit transaction.
Finally, we assume that the underlying lending platforms used by~\system are
secure. 

\section{Detailed zkSNARK-based System Construction}
\label{sec:detailedcontruction}
We now present a \zksnark-based construction of \system. 
\subsection{Building Blocks}
\label{sub:buildingnlocks}
\pparagraph{Hash Functions}
$H_{p}:\bset^* \rightarrow \FF$ is a preimage-resistant and
collision-resistant hash function that maps binary string to an element
in $\FF$, $H_{2p}:\FF\times \FF \rightarrow \FF$ be a collision-resistant hash
function that maps two elements in $\FF$ into an element in $\FF$.

\pparagraph{Deposit Commitments} A secure commitment scheme $(\pcom,
$ $\vcom)$ can be constructed using a secure hash function, $H_p:\bset^*\rightarrow \FF$, 
as follows: (1) $\pcom(m,r)$ returns $c=H_p(m||r)$, (2) $\vcom(c,m,r)$ verifies 
if $c \stackrel{?}{=} H_p(m||r)$.


In \system, before depositing into the contract, a client samples
 randomnesses, $k_{dep},r$ and computes the deposit commitment: $\cm
=H_p(k_{dep}||r)$ as a part of a deposit transaction.


\begin{figure}[b]
\centering
  \begin{tikzpicture}[level distance=.6cm,
    level 1/.style={sibling distance=2cm},
    level 2/.style={sibling distance=2cm},
    level 3/.style={sibling distance=2.3cm,level distance=.8cm},
    normal/.style={rectangle,draw,fill=black!20},
    root/.style={rectangle,draw}
    ]
    \node[root] (dep) at (0,0) {\scriptsize $\rt$}
    child {node {\scriptsize \dots}
      child { node[normal] {\scriptsize $H_{2p}(\cdot, \cdot)$} 
      child {node {\scriptsize $\cm_{dep,1}$} child { node[normal] {\scriptsize $H_p(\cdot)$} child {node {\scriptsize $k_1||r_1$}}}} 
      child {node {\scriptsize $\cm_{dep,2}$} child { node[normal] {\scriptsize $H_p(\cdot)$} child {node {\scriptsize $k_2||r_2$}}}}
      }
      child {node {\scriptsize \dots}}
    }
    child {node {\scriptsize \dots} 
        child {node {\scriptsize \dots}} 
        child {node[normal] {\scriptsize $H_{2p}(\cdot, \cdot)$} child {node {\scriptsize \dots}} child{ node {\scriptsize $0^n$}} 
        }};
    \end{tikzpicture}
    \caption{Illustrative example of the Merkle tree, $T_{dep}$. The tree keeps track of commitments
    from by clients' deposit transactions. The root of the tree, $\rt$ 
    is used to verify the NIZK proofs from withdrawing and redeeming-reward transactions.
    }
    \label{fig:trees}
\end{figure}
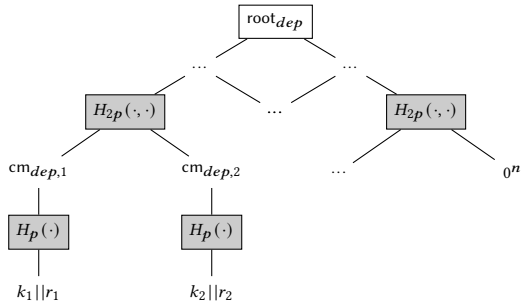

\pparagraph{Merkle Tree over Deposit Commitments, $T_{dep}$} 
The \system contract maintains a Merkle tree, $T_{dep}$, over all commitments.
a Merkle tree is an instance of an authenticated data structures for testing set membership~\cite{boneh-shoup-book-2020} (cf. \Cref{sec:preliminaries}).
The Merkle tree in the \system contract is a complete binary tree and initialized with zero values at its leaves.  
As deposit transactions arrive, the \system contract keeps track of the number of 
deposit transactions and updates the trees through the \acceptdeposit algorithm.
A Merkle tree can be constructed using a collision-resistant hash function,
$H_{2p}$.

We denote $\mpath_{i}$ the Merkle proof of $\cm_{i}$. 
We denote the Merkle tree root at block $h$ to be $\rt^{h}$.
We let $\rt.\mathsf{blockheight}$ to be the height of the
blockchain block when $\rt$ gets updated. 
\Cref{fig:trees} gives an illustrative example of the Merkle tree maintained 
by the \system contract.



\pparagraph{Withdrawal Proof} To withdraw coins from \system, a
  client needs to prove three conditions:  
\emph{(i)} the client knows the committed values of some existing commitments
used to compute the tree root via zkSnark proof, 
\emph{(ii)} the client did not withdaw in the past by passing a \emph{fresh} nullifier value, 
\emph{(iii)} the client knows the secret key used to issue the withdrawing transaction. 

The last condition prevents network adversaries from stealing a valid proof by
binding the public/private key to the zksnark proof.  
In particular, for a Merkle tree
$T$ with a root, $\rt$, a client needs issue a proof proving the following
relation:
\begin{equation}
\label{eq:relation}
\small
\begin{split}
R_{wdr}: 
\set{&\pk, \sn,  \rt; \gamechange{\text{$\sk, k_{dep}, r, \mpath_i$}}: \\
&\pk=\fun{extractPK}(\sk)\wedge\sn = H_p(k_{dep})\wedge~\\
&\cm = H_{p}(k_{dep}||r)\wedge
T.\mkverify(i,\cm, \rt, \mpath_{i}))}\\
\text{Where } &\pk,\sn, \rt \text{ are public values} \text{ and }\\ 
&\gamechange{$\sk, k_{dep},r,\mpath_{i}$}~\text{are private values.}  
\end{split}
\end{equation}
The nullifier value is used to ensure correctness by preventing clients from double-withdrawal.

\begin{figure}[t]
    \begin{pchstack}[boxed,center]
    \procedure[width=1em,linenumbering, bodylinesep=2pt]{$\fun{ContractSetUp}(\sparam)$}
    {
        \text{Sample } H_{p}:\bset^{*}\rightarrow \FF\text{ and } H_{2p}:\FF\times\FF \rightarrow \FF\\
        \text{Choose } \xcoin \in \mathbb{Z}_{>0} \text{ to be a fixed deposit amount}\\
        \text{Choose } \xcoin_{rwd} \in \mathbb{Z}_{>0} \text{ to be a fixed reward amount}\\
        \text{Choose } t_{con} \in \mathbb{Z}_{>0} \text{ to be condition for getting reward}\\
        \text{Choose } d \in \mathbb{Z}_{> 0}, \text{ Let } X=\set{x_1,\dots, x_{2^d}}  \\ 
        \t\t\t \text{ where } x_i = 0^{\lambda} \text{ for all } x_i \in X\\
        \text{Choose } \Sigma \text{ to be the lending platform}  \\ 
        {\text{Deploy } \Gamma_{\system} \text{ to be the interest distribution pool}}  \\ 
        \text{Initialize an empty tree }\rt = T.\initialize(\sparam, X), \\
        \text{Choose }k \in \mathbb{Z}_{>0},  
        \text{ set } \drootlist[i]=\rt, \pcskipln\\ 
        \t\t \text{for } 1\leq i \leq k\\
        \text{Set } \rtr^{\mathsf{curr}}=\rtr^{\mathsf{next}}=\rt, \idx = 1\\
        \text{Construct } C_{wdr} \text{ for relation described in~\Cref{eq:relation}}.\\
        \text{Let } \Pi \text{ be the \zksnark instance}. \pcskipln\\
        \t-\text{Run } (\ek_{dep},\vk_{dep})\leftarrow \Pi.\zksetup(\sparam, C_{wdr})\pcskipln\\
        \text{Initialize: } \mathsf{DepositList}=\set{}, \mathsf{NullifierList}=\set{},\\
        \text{Deploy smart contract \system  with parameters :} \pcskipln\\
        \pp = (\FF, H_{p}, H_{2p}, \xcoin, \xcoin_{rwd}, t_{con}, \Sigma, \Gamma_{\system} \pcskipln\\ 
        \t\t\t T, \idx, \drootlist, \rtr^{\mathsf{curr}}, \rtr^{\mathsf{next}},\pcskipln\\
        \t\t\t (\ek_{dep},\vk_{dep}),\mathsf{DepositList},\mathsf{NullifierList})
    }
    \end{pchstack}
    \caption{\system Setup. The public parameters, $\pp$, contains
    all information needed to interact with the $\system$ contract, and $\pp$ can be 
    queried by any client.
    }
    \label{fig:setup}
\end{figure}

\begin{figure*}[t]
  \procedureblock[bodylinesep=2pt]{Deposit Interactions}{
  \textbf{Client}(\sk, \xcoin) \<\< \textbf{\system Contract} \\ 
  \dbox{
  \begin{subprocedure}
  \procedure[linenumbering, bodylinesep=2pt]{$\CreateDTX(\sk, \xcoin):$}
    {
      \text{Sample } (k_{dep}, r) \sample \set{0,1}^{\lambda}\\ 
      \text{Compute } \cm=H_{p}(k_{dep}||r)\\
      \pcreturn \txd=(\xcoin, \cm) \text{ and } \witness=(k_{dep}, r)
    }
  \end{subprocedure}
  }\<\sendmessageright*{\txd}\<
  \dbox{
    \begin{subprocedure}
      \procedure[linenumbering, bodylinesep=2pt]{$\acceptdeposit(\txd)$}{
        \text{Parse } \txd = (\xcoin', \cm)\\
        \text{Require } \xcoin {=} \xcoin' \text{ and } \idx < 2^d\pcskipln\\
        \text{/* Invest to the lending platform */}\\
        \text{Execute } \Sigma.\fun{Deposit}(\xcoin) \text{ to obtain } \xcoin_\Sigma\\
        \text{Append } \cm \text{ to } \dpl\\
        \text{Increment } \idx=\idx+1\\
        \text{Compute } \pcskipln \\ 
        \text{ - }\mathsf{root}_{new} = T_{dep}.\update(\idx, \cm, \dpl)\\
        \text{Append } \rt \text{ to } \drootlist\pcskipln\\
        \text{/* Update reward roots*/}\\
        \text{If } \mathsf{Block.Height} - \nextrroot.\mathsf{blockheight} \geq t_{con}: \pcskipln\\
        \text{ - Set } \currentrroot = \nextrroot \pcskipln\\
        \text{ - Set } \nextrroot = \mathsf{root}_{new}\\
        \pcreturn 1
      }
    \end{subprocedure}
    }\\
  }
  \caption{\system's deposit interactions between the client (Client's \CreateDTX algorithm) and \system contract (\system's \acceptdeposit algorithm). 
  Transaction $\txd$ is signed by $\sk$. $\mathsf{Block.Height}$ denotes the block height of the block containing $\txd$}
  \label{fig:dtx}
\end{figure*}

\pparagraph{Reward Proof} 
Intuitively, to prove that funds remained in the system for a certain time period, 
users can simply prove to the contract that they know some commitment their, $\cm$, that
is a member of an older Merkle root.
To achieve such condition, the \system contract always maintains an $t_{con}$-blocks-old
Merkle root that serves as an anchor for clients to issue the reward proof.
Similar to withdrawing, to redeem, clients need to nullify
the old commitment, $\cm$, by issuing a nullifier value, $\sn$, 
and submit a new commitment, $\cm'$ to be eligible for future redeems and withdrawals. 
This requirement allows AMR to maintain system correctness and hide the link
between reward-redeeming and withdrawing transactions. 

In summary, to redeem coins from \system, a client needs to prove that: 
\emph{(i)} the client knows the committed value of some existing commitments used
to compute the current reward Merkle tree root via a zkSnark proof,
\emph{(ii)} the client did not withdraw in the past by passing a \emph{fresh}
nullifier value \sn, and
\emph{(iii)} Finally, the client needs to refresh its original deposit by
submitting a new commitment to be eligible for future reward redemptions and
withdrawals.

\subsection{Contract Setup}
\label{sub:contractsetup}

Let $\FF$ be the finite field used in \system,
during the \system contract setup phase, 
the setup algorithm
samples secure hash functions $H_p:\bset^* \rightarrow \FF, H_{2p}:\FF\times
\FF \rightarrow \FF$ from secure collision-resistant hash families.
The \system contract is initialized with several parameters: 
$\xcoin$ for the fixed amount of coins to be mixed,
$\xcoin_{rwd}$ indicating the fixed amount of coins to be rewarded,
and $t_{con}$ specifying the minimum number of blocks that clients need to wait before redeeming rewards.

\pparagraph{Setting up Merkle Trees}
Let $T$ be the Merkle tree of depth $d$, the setup algorithm described in~\cref{sub:csetup} 
initializes $T$ with zero leaves and initializes $\idx=1$ to keep track of latest
deposits.
Also, the algorithm initializes two lists: 
$\drootlist$ to be the list of $k$ most recent roots of $T$.
Finally, the contract keeps track of the current reward root, $\rtr^{\mathsf{curr}}$ 
that is used by clients to form reward proofs. 
and the next reward root $\rtr^{\mathsf{next}}$. 
Recall that $t_{con}$ to be the minimum number of blocks that clients need to wait
before redeeming a reward, we require:
$\rtr^{\mathsf{next}}.\mathsf{blockheight}-\rtr^{\mathsf{curr}}.\mathsf{blockheight}\geq t_{con}$.
This approach helps the \system contract maintain $t_{con}$-blocks-old reward root without storing all other roots.

\pparagraph{Setting up \zksnark parameters}
Let $\Pi$ be the \zksnark instance used in \system, the setup algorithm \Cref{sub:csetup} 
constructs circuit $C_{wdr}$ capturing the relation described in~\Cref{eq:relation}.
Then, the setup algorithm runs $\Pi.\zksetup$ on the circuit to obtain two keys, $(\ek_{dep},\vk_{dep})$.

\pparagraph{Setting up commitments and nullifier lists} 
The \system contract
is initialized with two empty lists: a list, $\dpl$, that contains all
$\cm$ included in depositing and reward-redeeming transactions,
a list, $\textsf{NullifierList}$, that contains all unique identifiers (i.e.
$\sn$) appeared in withdrawing and reward-redeeming transactions.
\Cref{fig:setup} formally describes this setup algorithm.


\subsection{Client Algorithms}
These following algorithms specify how clients interact with the \system smart contract.

\begin{figure*}[t]
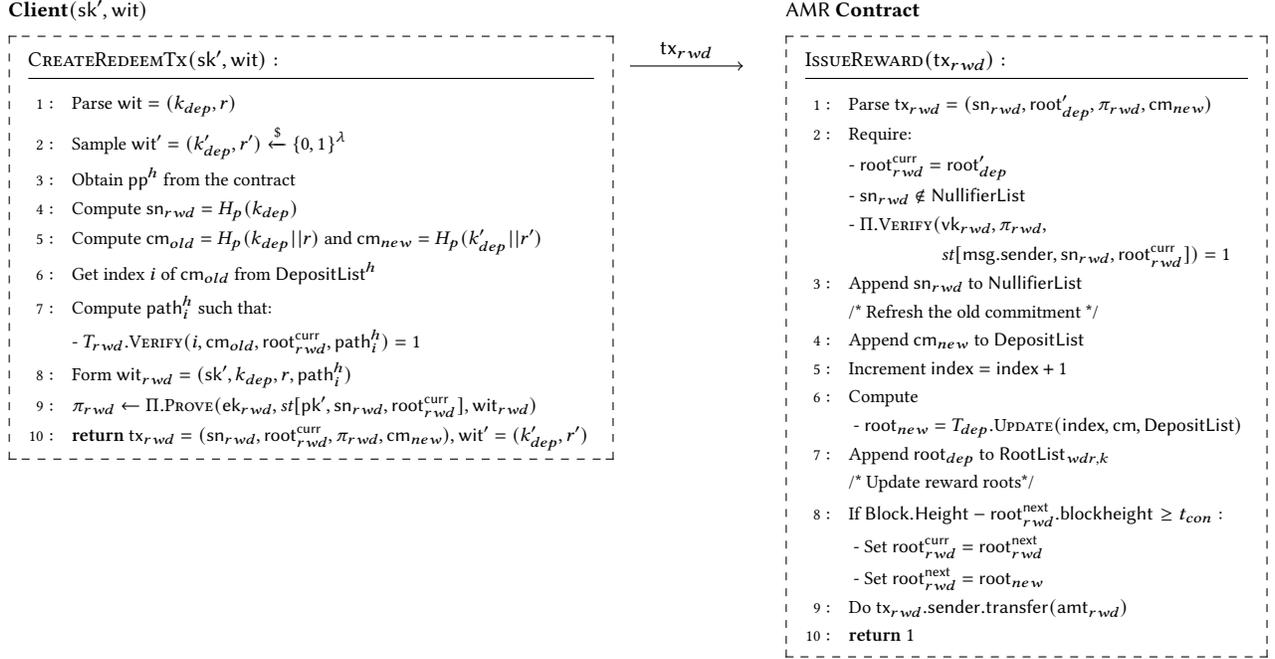

  \procedureblock[bodylinesep=3pt]{Reward-Redeeming Interactions}{
  \textbf{Client}(\sk', \witness) \<\< \textbf{\system Contract} \\ 
  \dbox{
   \begin{subprocedure}
      \procedure[linenumbering, bodylinesep=3pt]{$\CreateRTX(\sk', \witness):$}
      {
      \text{Parse } \witness = (k_{dep}, r)\\
      \text{Sample } \witness' = (k'_{dep}, r') \sample \bset^\lambda \\ 
      \text{Obtain } \pp^h \text{ from the contract} \\
      \text{Compute } \sn_{rwd} = H_p(k_{dep})\\
      \text{Compute } \cm_{old} = H_p(k_{dep}||r) \text{ and } \cm_{new} = H_p(k'_{dep}||r')\\
      \text{Get index } i \text{ of } \cm_{old} \text{ from } \dpl^h\\
      \text{Compute } \mpath^h_i \text{ such  that: }  \pcskipln\\
      \text{- } T_{rwd}.\mkverify(i,\cm_{old}, \currentrroot, \mpath^h_i) = 1\\
      \text{Form } \witness_{rwd} = (\sk', k_{dep}, r, \mpath_{i}^h)\\
      \pi_{rwd} \leftarrow \Pi.\fun{Prove}(\ek_{rwd}, \statement[ \pk', \mathsf{sn}_{rwd}, \currentrroot], \witness_{rwd})\\
      \pcreturn \txreward=(\mathsf{sn}_{rwd}, \rtr^{\mathsf{curr}}, \pi_{rwd}, \cm_{new}), \witness'=(k'_{dep}, r')
      }
    \end{subprocedure}
  }\<\sendmessageright{length=1.5cm,top=$\txreward$}\<
  \dbox{
    \begin{subprocedure}
      \procedure[linenumbering, bodylinesep=3pt]{$\issuereward(\txreward):$}
    {
        \text{Parse } \txreward = (\sn_{rwd}, \rt', \pi_{rwd}, \cm_{new})\\
        \text{Require: } \pcskipln\\
        \text{- } \currentrroot = \rt' \pcskipln\\
        \text{- } \sn_{rwd} \notin \rwnl\pcskipln \\
        \text{- } \Pi.\zkverify(\vk_{rwd},\pi_{rwd}, \pcskipln\\
        \t\t\t\t\t\statement[\mathsf{msg.sender}, \sn_{rwd}, \currentrroot]) = 1\\
        \text{Append } \sn_{rwd} \text{ to } \rwnl \pcskipln\\
        \text{/* Refresh the old commitment */}\\
        \text{Append } \cm_{new} \text{ to } \dpl\\
        \text{Increment } \idx=\idx+1\\
        \text{Compute } \pcskipln \\ 
        \text{ - }\mathsf{root}_{new} = T_{dep}.\update(\idx, \cm, \dpl)\\
        \text{Append } \rt \text{ to } \drootlist\pcskipln\\
        \text{/* Update reward roots*/}\\
        \text{If } \mathsf{Block.Height} - \nextrroot.\mathsf{blockheight} \geq t_{con}: \pcskipln\\
        \text{ - Set } \currentrroot = \nextrroot \pcskipln\\
        \text{ - Set } \nextrroot = \mathsf{root}_{new}\\
        \text{Do } \txreward\mathsf{.sender.transfer}(\xcoin_{rwd})\\
        \pcreturn 1
    }
    \end{subprocedure}
    }\\
  }
  \caption{\system's reward-redeeming interactions between the client (Client's \CreateRTX algorithm) and \system contract (\system's \issuereward algorithm). 
  $\pp^h$ denotes the state of the contract at block height $h$. 
  The reward-redeeming transaction, $\mathsf{tx}_{rwd}$, contains the proof $\pi_{rwd}$
  that proves to the $\mathsf{AMR}$ contract the client's knowledge of a $t_{con}$-blocks
  old deposit, $\cm_{old}= H_p(k_{dep}||r)$ which is a valid member of the Merkle
  tree with the root, $\currentrroot$.
  $\sn_{rwd}$ is used to nullify the old commitment, and $\cm_{new}$
  is used to refresh the old commitment. 
  $\txreward$ is signed by $\sk'$.
  $\mathsf{Block.Height}$ denotes the block height of the block containing transactions $\txr$}
  \label{fig:rtx}
\end{figure*}

\pparagraph{Depositing} \CreateDTX allows a client to deposit coins into the contract 
and outputs secret notes, $\witness$, that can later be used to withdraw coins
or obtain a reward.  

\begin{figure*}[t]
  \procedureblock[bodylinesep=3pt]{Withdraw Interactions}{
  \textbf{Client}(\sk', \witness) \<\< \textbf{\system Contract} \\ 
  \dbox{
   \begin{subprocedure}
    \procedure[linenumbering, bodylinesep=2pt]{$\CreateWTX(\sk', \witness):$}
    {
    \text{Parse } \witness = (k_{dep}, r)\\
    \text{Obtain } \pp^h \text{ from the contract}\\
    \text{Compute } \sn_{wdr} = H_p(k_{dep})\\
    \text{Compute } \cm = H_p(k_{dep}||r)\\
    \text{Get index $i$ of } \cm \text{ from } \dpl^h\\
    \text{Choose } \rt \in \drootlist \\
    \text{Compute } \mpath^h_{dep,i}\text{ such that } \pcskipln \\
    \text{ - } T.\mkverify(i,\cm, \rt, \mpath^h_{dep,i}) = 1 \\
    \text{Form: } \witness_{dep} = (\sk', k_{dep}, r, \mpath^h_{dep,i})\\
    \pi_{wdr} \leftarrow \Pi.\fun{Prove}(\ek_{dep}, \statement[\pk',\sn_{wdr},\rt],\witness_{dep})\\
    \pcreturn \txw = (\sn_{wdr},\rt,\pi_{wdr})
    }
    \end{subprocedure}
  }\<\sendmessageright{length=1.5cm,top=$\txw$}\<
  \dbox{
    \begin{subprocedure}
       \procedure[linenumbering, bodylinesep=3pt]{$\issuewithdraw(\txw):$}
    {
        \text{Parse } \txw = (\sn_{wdr},\rt,\pi_{wdr})\\
        \text{Require } \pcskipln\\
        \text{- } \rt \in \drootlist \pcskipln\\
        \text{- } \sn_{wdr} \notin \wnl   \pcskipln\\
        \text{- } \Pi.\zkverify(\vk_{dep},\pi_{wdr}, \pcskipln\\ 
        \t\t\t\t\t\statement[\mathsf{msg.sender}, \sn_{wdr}, \sn_{rwd},\rt]) = 1\\
        \text{Let } \mathsf{balance}_{\Sigma}\text{ be the } \Sigma \text{ token balance}\\
        \text{Let } \mathsf{NoDepositsLeft} \text{ be the number of deposits remained}\pcskipln\\
        \text{/* Redeeming from the lending platform */}\\
        \text{Redeem } \mathsf{balance}+R\text{ by executing } \Sigma.\fun{Redeem}(\mathsf{balance}_{\Sigma})\pcskipln\\
        \t\t \text{ where } \mathsf{balance}=\key{NoDepositsLeft}\cdot \xcoin\\
        \text{Append } \sn_{wdr}\text{ to } \wnl \pcskipln\\
        \text{/* Send the accrued interest to the distribution pool */}\\
        \text{Do } \Gamma_{\system}.\mathsf{transfer}({R}/{\mathsf{NoDepositsLeft}})\pcskipln\\
        \text{/* Send the original deposit to sender*/}\\
        \text{Do } \texttt{tx}_{wdr}\mathsf{.sender.transfer}(\xcoin)\pcskipln\\
        \text{/* Reinvest into the lending platform */}\\
        \text{Let } \mathsf{removedFund}=\xcoin+ {R}/{\mathsf{NoDepositsLeft}}\\
        \text{Execute } \Sigma.\fun{Deposit}(\mathsf{balance}+R-\mathsf{removedFund})\\
        \pcreturn 1
    }
    \end{subprocedure}
    }\\
  }
  \caption{
  \system's deposit interactions between the client (Client's \CreateWTX algorithm) 
  and \system contract (\system's \issuewithdraw algorithm). 
  $\pp^h$ denotes the state of the contract at block height $h$.
  The withdrawing transaction, $\tx_{wdr}$, contains the proof $\pi_{wdr}$, that
  proves to the $\system$ contract the client's knowledge of a commitment, $\cm$,
  which is a valid member of the Merkle tree with the root, $\rt$. 
  $\sn_{wdr}$ is used to nullify the old commitment, $\cm$.
  Finally, $\txw$ is signed by $\sk'$. 
  }
  \label{fig:wtx}
\end{figure*}


\pparagraph{Redeeming Reward} \CreateRTX allows clients with the secret note, $\witness$, 
and the secret key $\sk'$ to issue a proof, $\pi_{rwd}$, to prove to the \system contract
that that client has not withdrawn their deposited coins after certain number of block counts.
In order to form such NIZK proof, the client obtains the current state of the contract
to compute private inputs (i.e.\ Merkle path) for the \zksnark proof generation. 
Also, the proof generation requires the client to use an older root maintained by the contract
as part of the computation. 
This approach allows a client to prove to the contract that the client's transaction was deposited 
before the root was computed. 
Along with the NIZK proof, a client will include the nullifier value as part of the transaction to prevent 
double-redemption from the \system contract. 
Finally, the client includes a new commitment value, $\cm'$, in the reward-redeeming 
transaction 
to be eligible for future withdrawal and reward-redemption.


\pparagraph{Withdrawing} \CreateWTX allows a client with the secret note, $\witness$, 
and secret key $\sk'$ to issue proofs, $\pi_{wdr}$, to withdraw $\xcoin$ to the public key
$\pk'$. In this step, \system requires the client to issue a proof to verify that
the client has deposited coins in the past along with a nullifier value $\sn$
to prove that those coins have not been withdrawn before 
and to prevent clients 
from withdrawing coins without having any coins deposited to the \system contract. 

\subsection{Contract Algorithms}
In this part, we formally define the contract algorithms.

\pparagraph{Accepting Deposit} 
Upon receiving a deposit transaction from an externally owned account, the
contract verifies the amount $\xcoin$, updates the tree structure, recomputes
the Merkle roots for the Merkle tree, and updates the $\dpl$ list.
Next, the \system contract deposits $\xcoin$ into the lending platform $\Sigma$
to retrieve $\xcoin_\Sigma$. 
Finally, depending on the number of blocks mined,
the contract always maintains a $t_{con}$-blocks-old root, so that
clients use it to redeem rewards.
\Cref{fig:dtx} formally describes this procedure. 



\pparagraph{Issuing Reward} 
Upon receiving reward transactions, the contract 
verifies that the proof, $\pi_{rwd}$, is valid with the $\currentrroot$, 
and the nullifier $\sn_{rwd}$ is not in the $\rwnl$. 
Once the verification passes, the contract updates the $\rwnl$ to prevent future double-redemption
and double-withdrawal. 
Here, we note that $\currentrroot$ is the old state 
of the reward tree; therefore, being able to prove the membership of this root, 
one can prove that their deposit has not been withdrawn.
Finally, the $\system$ contract updates the Merkle tree with the new commitment,
$\cm_{new}$. This update is similar to the deposit phase, and it helps the client
to refresh their original commitment to be eligible for future redemption and withdrawals.

\pparagraph{Issuing Withdraw}
Upon receiving withdrawal transaction, the contract verifies the proof,
$\pi_{wdr}$ and verifies that the nullifier $\sn_{wdr}$ is not
in the $\wnl$. 
To prevent future double-withdrawal, the \system contract then appends 
$\sn_{wdr}$ to $\rwnl$.
Then, the \system contract redeems all deposited funds
from lending platform $\Sigma$ along with the accrued interest $R$. 
Finally, the \system deposits $\mathsf{\xcoin}$ to the
user's address and redeposits leftover funds into lending platforms.
To avoid leakage in distributing accrued interest, $\system$ deposits the interest into a
separate pool, $\Gamma_{\system}$. 
Only users who hold the governance tokens obtained from rewards can later obtain this
interest. 

\Cref{fig:dtx}, \Cref{fig:rtx}, and \Cref{fig:wtx} formally describe the
interactions between the clients and the smart contract in \system.

\pparagraph{Distributing Accrued Interest}
We adapt the time-weight voting proposed by Curve~\cite{curve} for this distribution step.
In particular, in \system, the pool, $\Gamma_{\system}$, receives a portion of the total accrued interest upon each withdrawal.
Clients need to lock governance tokens to the pool to be eligible for
redeeming this interest. The \system pool periodically distributes the total 
accrued interest proportionally to clients based on their voting power. 


Client's voting power is calculated based on their amount of governance tokens and how long they are willing to lock those tokens in the distribution pool. 
The pool requires two main functionalities: \fun{CreateLock} and \fun{Claim}. 
The pool is initialized with a value $t_{max}$ denoting
the maximum amount of blocks that client can lock their governance tokens. 

\begin{itemize}[leftmargin=*]
  \item \fun{CreateLock}$(\tx_{lock})$ takes as input
    the locking transaction, $\tx_{lock}$, from the client. The locking transaction contains 
    the governance tokens, $\amt_{rwd}$, and $t_{lock}$ specify the number of blocks that the client will lock
    $\amt_{rwd}$ in the pool. At any given point in time, the voting power of the client is $\amt_{rwd}\cdot \frac{t}{t_{max}}$ 
    where $t$ is the time left to unlock $t\leq t_{lock}$.
  \item \fun{Claim}() is a contract function that can be periodically triggered by the clients. 
  When a client triggers this function, the pool calculates the client's current voting weight as $w = \amt_{rwd}\cdot \frac{t}{t_{max}}$, 
  and the total voting power of all users, $W$. Finally, the pool distributes the accrued interest proportionally to the client according their voting power and the total voting power, $\frac{w}{W}$.
\end{itemize}

The main goal of time-weighted voting is to distribute more reward not only to users 
who \emph{contribute} more to \system (i.e., having more governance tokens) 
but also to users who \emph{commit} more to the system (i.e., locking their stakes for a longer period).
Finally, for a more detailed description of this time-weight voting technique, we refer interested readers to~\cite{curve}.

\section{System Analysis}
\label{sec:systemanalysis}
In this section, we informally discuss how \system achieves the security goals
mentioned in~\Cref{sub:systemgoals}. 
As mentioned in \Cref{sub:threatmodels}, the underlying cryptographic
primitives (i.e., \zksnark, commitment scheme, hash functions) are assumed to be secure,
and \system's depositing and withdrawing functionalities can be thought as the 
shielding and de-shielding transactions in ZCash with a fixed denomination.
Therefore, the security of \system follows from the security of \zksnark-based 
applications like ZCash~\cite{sasson2014zerocash}. 
In particular, the malicious outsider will not be able to learn any information
from the public data. 
However, adversaries can still guess the pair-wise link between a withdrawing,
a depositing, and reward-redeeming transactions. 
The probability of guessing correctly largely depends on the number of
deposits, redemptions, and withdrawals issued to \system.
Thus, we need to understand this
adversarial probability to quantify the privacy offered by \system. 


\subsection{Privacy Metric}
\label{sub:metric}

We let $h$ be the height of the blockchain,  we define:
     $\anoset$ be the set of commitments issued to the \system contract until block height $h$ by \emph{honest} users, and the adversary does not know
     the preimages of those commitments.
     $\mathsf{NullifierSet}^h$ be the set of nullifiers appeared in either 
     reward-redeeming or withdrawing transactions issued to the \system until block height $h$ by \emph{honest} users.
$\mathsf{AnomSet}$ and $\mathsf{NullifierSet}$ are always available to the adversary. 
We also assume that $|\anoset| - |\mathsf{NullifierSet}^h| > 0$ for all $h$. 

We say $\cm$ originates $\sn$, when $k_{dep}$ is used to compute both $\cm$ in $\txd$ and $\sn_{wdr}$ in $\txw$ or $\txreward$.
We define:
\begin{itemize}[leftmargin=*]
    \item $\cm \stackrel{link}{\leftarrow}\sn :$ if the value $k$
      used to compute $\cm = H_{p}(\textcolor{blue}{k}||r) \in \txd$ or $\in \txreward$ is
      equal to the value $k$ used to compute $\sn =
      H_p(\textcolor{blue}{k}) \in \txreward$ or $\in \txw$.
\end{itemize}

{
  The reward linking advantage is the probability that an adversary can
  output the correct commitments that originates the nullifier value appeared in reward-redeeming transactions.
  We define that probability is as follow:
}

\begin{definition} 
\label{def:rdlinking}
(Reward Linking Advantage) 
Let $\mathcal A$ be the PPT adversary, 
$\txreward^{h+1}$ be the only valid reward-redeeming transaction issued at block $h+1$ from an honest user. Let $\sn^{h+1}_{rwd}$ be the nullifier appeared in $\txreward^{h+1}$ 
We define the adversarial advantage as follow:
    $$\mathsf{Adv}_{\mathcal A,rwd}^{h}=\Pr[\mathcal{A}(\txreward^{h+1}) \rightarrow \cm \in \anoset \text{ s.t. } \cm \stackrel{link}{\leftarrow}\sn_{wdr}^{h+1}]$$
\end{definition}

Similarly, the adversarial advantage in linking withdrawing transaction to
other transactions is the probability that an adversary can guess correctly the
commitment that originates the nullifier value appeared in withdrawing
transaction.  

\begin{definition} 
\label{def:wdlinking}
(Withdraw Linking Advantage) Let $\mathcal A$ be the PPT adversary, 
$\txw^{h+1}$ be the only valid withdrawing transaction issued at block $h+1$ from an honest user. 
Let $\sn^{h+1}_{wdr}$ be the nullifier appeared in $\txw^{h+1}$. 
We define the adversarial advantage as follow:
    $$\mathsf{Adv}_{\mathcal A,wdr}^{h}=\Pr[\mathcal{A}(\txw^{h+1}) \rightarrow \cm \in \anoset \text{ s.t. } \cm \stackrel{link}{\leftarrow}\sn_{wdr}^{h+1} ]$$
\end{definition}


We assume that the deposit addresses are independent and unlinkable accounts for our privacy metric to hold.  If the same entity deposits from different addresses, but a blockchain analysis allows to link those addresses, the anonymity set would only grow by at most 1 deposit.

\subsection{Privacy Analysis}

\pparagraph{Systems without reward} 
In a vanilla \system system that only supports depositing and withdrawing functionalities,
a withdrawal transaction can be at the origin of any deposit transactions of honest users
before the withdrawal transaction
, under the assumption that all underlying cryptographic primitives are secure. 
The adversarial advantage in linking withdrawing transaction to the original deposit transaction is: 
$\mathsf{Adv}_{\mathcal A,wdr}^{h} = 1/|\depset|+\mathsf{negl}(\lambda)$
where $\mathsf{negl}(\lambda)$ is the adversarial advantage in breaking the underlying
cryptographic primitive. 

\pparagraph{System with reward} Because \system involves redeeming transactions, we need to analyze the adversarial advantages under different scenarios. 
In the following, we show the adversarial advantage in linking different transactions through the following claims.  
\begin{claim}
Assuming that all underlying cryptographic primitives are secure, the adversarial
advantage in linking reward-redeeming transaction to other transactions as defined in \Cref{def:rdlinking}
is less than ${1}/{|\mathsf{AnomSet}^{h-t_{con}}|} + \mathsf{negl}(\lambda)$
\end{claim}
\textit{Sketch.} 
\system is parameterized with
the value $t_{con}$, the number of blocks that a client needs to wait to 
be eligible for a reward.
The adversary observes
a redeeming transaction issued to the \system contract after block height $h+1$ from an honest user.
A valid redeeming transaction indicates to the adversary that the sender has issued commitment into the system at least
$h-t_{con}$ blocks ago. 
Therefore, the probability that the adversary links the redeeming transaction to the correct commitment is hence: 
$\mathsf{Adv}_{\mathcal A, rwd}^{h} \leq {1}/{|\mathsf{AnomSet}^{h-t_{con}}|}+\mathsf{negl}(\lambda)$ where $\mathsf{negl}(\lambda)$ is the adversarial advantage in breaking the underlying cryptographic primitive.  

\begin{claim}
Assuming that all underlying cryptographic primitives are secure, the adversarial
advantage in linking between withdrawing transaction to other transactions as defined in \Cref{def:wdlinking}
is ${1}/|\anoset| + \mathsf{negl}(\lambda)$.
\end{claim}
\textit{Sketch.} 
Since we assume that the underlying cryptographic primitives are secure, the adversarial advantage in guessing correctly by breaking those primitives is negligible. Moreover, since each deposit and reward-redeeming transaction in
\system adds another leaf to the Merkle tree, the probability of guessing a correct 
leaf is equal to the number of Merkle leaves that are not controlled by the adversary.
In another word, the probability is ${1}/(|\anoset|)$. 
Therefore, the adversarial advantage, $\mathsf{Adv}_{\mathcal A, wdr}^{h} \leq {1}/|\anoset| +\mathsf{negl}(\lambda)$ where $\mathsf{negl}(\lambda)$ is the adversarial advantage in breaking the underlying cryptographic primitive.

{ 
In summary, in \system, given a reward-redeeming transaction, to guess the correct commitment, the adversary can reduce the size of the anonymity set by narrowing the search window to $t_{con}$ blocks before the block containing the reward-redeeming transaction.
}
On the other hand, given a withdrawing transaction, the adversary's advantage in guessing the correct commitment is still the same as the adversarial advantage in the system without reward, $1/|\depset|+\mathsf{negl}(\lambda)$. 
The main reason is that, in \system, beside each deposits, each reward-redeeming transaction also adds one additional commitment to 
the Anonymity Set, $\mathsf{AnomSet}$.
Therefore, \system offers a bigger anonymity set than system without reward.

\pparagraph{Privacy of the Accrued Interest Distribution}
One na\"{i}ve way to distribute the accrued interest is to split the total accrued interest equally among depositors. This approach reveals nothing about
the original deposits. However, it introduces an unfair allocation of interest as users joining the system later receive the same amount of interest as users joining the system earlier. 

In \system, to achieve fairness in the accrued interest distribution, 
\system allows users with governance tokens to lock their tokens in a separated pool (i.e., $\Gamma_{\system}$).
This pool receives a portion of the accrued interest from the \system contract upon each withdrawal, and it periodically distributes the total accrued interest to addresses that lock their governance token into the pool. 
The amount each address receives is based on the number of governance tokens and how long those tokens are locked. 
{
It's not difficult to see that 
\system ensures the fairness in the accrued interest allocation because 
only users contributing more to the anonymity set of \system, can redeem more governance
tokens; therefore, they can obtain more accrued interest.  
}



\subsection{Other goals achieved by \system}
In addition to the privacy goal, we briefly explain how \system achieves the other goals 
defined in~\Cref{sub:systemgoals}.

\pparagraph{Correctness} \system satisfies correctness. If an adversary can provide a withdrawal transaction that verifies without depositing any coins into the system,
there are two possible scenarios: First, 
the adversary can derive a new valid transaction for the current state of the contract 
(i.e. observing commitment list), 
or it intercepts a withdrawal transaction and replaces the recipient address with
its address. However, in the first case, it implies that the adversary breaks
the preimage-resistant security of the underlying hash function $H_p(\cdot)$, and the second case implies
that the adversary breaks the security of the \zksnark instance.

\pparagraph{Availability} We argue that \system satisfies availability. Unlike
existing centralized tumbler designs~\cite{heilman2017tumblebit}, the availability
of the system relies on the fact that the tumbler has to stay online.
Similar to M\"{o}bius~\cite{meiklejohn2018mobius}, \system is a smart contract that executes autonomously on the blockchain, 
so adversary cannot prevent clients from interacting (i.e.,\ reading and writing) with the blockchain.

\pparagraph{Front-Running Resilience} Recall that the \system contract stores 
a list of $k$ recent roots. To invalidate a withdrawal transaction, an adversary needs to ``front-run'' at least $k$ deposit transactions before a withdrawal transaction. 
Thus, one can choose the value $k$ to be sufficiently large so that the cost of 
attacking is too expensive for the adversary to carry out. 
More specifically, to invalidate a single deposit transaction, 
the amount of token an adversary needs to have are at least $k\times (\xcoin + \mathsf{fee}_{dep})$
where $\xcoin$ is the fixed denomination specified in~\cref{sub:contractsetup} and $\mathsf{fee}_{dep}$ is the deposit fee.  
For example, if we set $k=1000$, $\xcoin = 10$, assuming $\mathsf{fee}_{dep}=0.02$, and let the token be \textit{ether}, the adversary needs at least $k\times(\xcoin+\mathsf{fee}_{dep}) = 10,020$ \textit{ethers} ($38m$ USD) to carry out the attack, and the adversary will lose at least $20$ \textit{ethers} ($76,000$ USD) in term of fee.

%

\section{Evaluation}
\label{sec:evaluation}

\begin{table}[t]
\resizebox{\columnwidth}{!}{%
\begin{tabular}{@{}ccccccc@{}}
\toprule
  \multirow{1}{*}{\textbf{Tree Depth}} &
  \multicolumn{2}{c}{\textbf{\# Constraints}} &
  \multicolumn{2}{c}{\textbf{Setup Time}} &
  \multicolumn{2}{c}{\textbf{Keys Size}} \\ 
  \cmidrule(l{5pt}r{5pt}){2-3}  \cmidrule(l{5pt}r{5pt}){4-5}  \cmidrule(l{5pt}r{5pt}){6-7}
  &
  \multicolumn{2}{c}{$C_{wdr}$} &
  \multicolumn{2}{c}{$t_{wdr}$} &
  \multicolumn{2}{c}{$(\ek_{wdr}, \vk_{wdr} = 640 B)$} \\ 
  \cmidrule(l{5pt}r{5pt}){2-3}  \cmidrule(l{5pt}r{5pt}){4-5}  \cmidrule(l{5pt}r{5pt}){6-7}
   & Poseidon & MiMC     & Poseidon  & MiMC      & Poseidon & MiMC     \\ \midrule
10 & $4,245$  & $15,045$ & $86.56s$ & $246.99s$  & $4.3$MB  & $7.3$MB \\
15 & $5,460$  & $21,660$ & $107.34s$ & $377.44s$ & $5.8$MB &  $11.1$MB \\
20 & $6,675$  & $28,275$ & $126.51s$ & $465.27s$ & $7.3$MB &  $13.8$MB \\
25 & $7,890$  & $34,890$ & $146.41s$ & $642.04s$ & $8.8$MB &  $18.6$MB \\
30 & $9,105$  & $41,505$ & $185.64s$ & $729.03s$ & $10.8$MB &  $21.4$MB \\ \bottomrule
\end{tabular}%
}
\caption{zk-SNARK Setup Cost}
\label{tab:trusted-setup}
\end{table}
\subsection{Parameters}
\pparagraph{Choice of cryptographic primitives} We use Groth's
zkSNARK~\cite{groth-zksnark-2016} as our instance of \zksnark due to its
efficiency in term of proofs' size and verifier's computations.  
For cryptographic hash functions, 
we use a Pedersen hash
function~\cite{pedersen-hash} for $H_{p}$ and evaluate \system using two
different choices of hash functions for $H_{2p}$: the MiMC ~\cite{MiMC-hash}
and the Poseidon hash function~\cite{poseidon-hash-2019}. 
Arithmetic circuits using MiMC and Poseidon hash yield a lower number of
constraints and operations when compared to arithmetic circuits relying on
other hash functions~\cite{inefficency-sha256,MiMC-hash} (i.e. SHA-256,
Keccak).
Moreover, 
both MiMC and Poseidon hash functions are not only designed specifically for
SNARK applications, but also highly efficient for Ethereum smart contract
applications in terms of gas costs. 
Finally, as discussed in \Cref{sub:buildingnlocks},
the commitment scheme and the Merkle tree can be directly instantiated using
Pedersen and MiMC/Poseidon hash functions.
\begin{figure}[b]
    \centering
    \begin{minipage}{\columnwidth}
    \centering
    \includegraphics[width=\columnwidth]{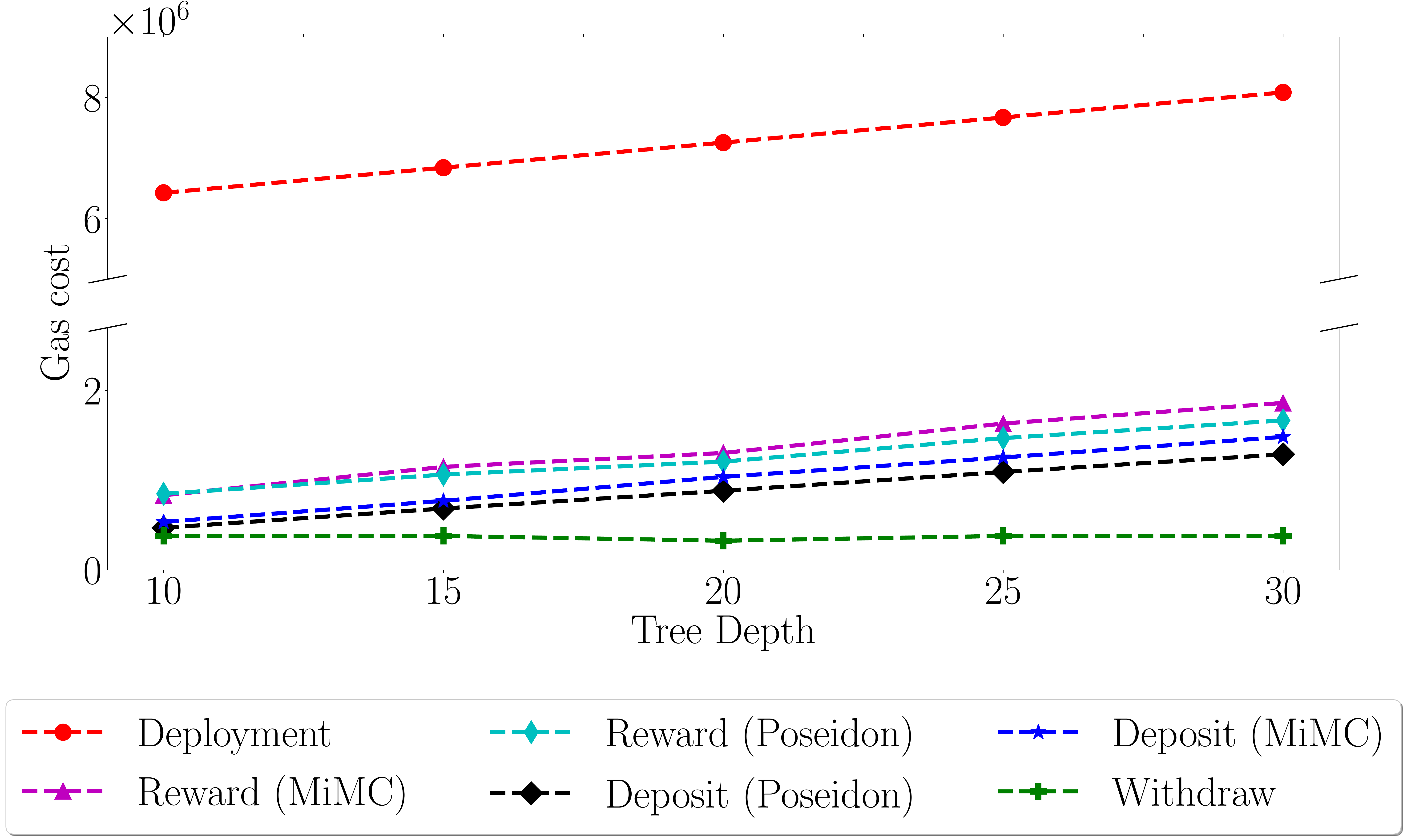}
    \caption{On-chain Costs of Deployments, Deposit, Withdrawal, and Reward
    Redemption for Different Tree Depths and Hash Functions.
    }
    \label{fig:onchain-cost}
    \end{minipage}
\end{figure}

\pparagraph{Software} 
For the arithmetic circuit construction, we use the \texttt{Circom}
library~\cite{circom-lib} to construct the withdrawing circuit, $C_{wdr}$ for
the relation described in \Cref{eq:relation}.
We use Groth's \zksnark proof system implemented by the \texttt{snarkjs}
library~\cite{snarkjs-lib} to develop the client's algorithms (cf.\
\Cref{sub:clientalgorithms}), 
and to perform the trusted setup for obtaining the proving and evaluation keys
for the \system contract and clients. We deploy \system to the Ethereum Kovan
testnet~\footnote{\system's address:
0xdE992c4fBd0f39E5c0356e6365Bcfafa1e94970b}~\footnote{A demo video \system can
be found at the following URL: \url{https://youtu.be/-oAQlsRTF08}}. 
\system contract consists 1013 lines of Solidity code.

\pparagraph{Hardware} We conducted our experiment on a commodity desktop machine,
which is equipped with an Intel Core i$5$-$7400$ @3.800GHz CPU, 32GB RAM.

\subsection{Performance}
We measure the performance and the cost of \system using the following tree depths 
$d=10,15,20,25,30$. 

\pparagraph{\zksnark Setup}
\Cref{tab:trusted-setup} presents an overall performance of the \zksnark setup
for the withdraw circuit. For the MiMC hash function, for a tree of depth $d$,
the withdraw circuit has $1,815+1,323\times d$ constraints.
For the Poseidon hash function, the withdraw circuit has $1815+243\times h$ constraints.
\begin{figure}[t]
    \begin{minipage}{\columnwidth}
    \centering
    \includegraphics[width=\columnwidth]{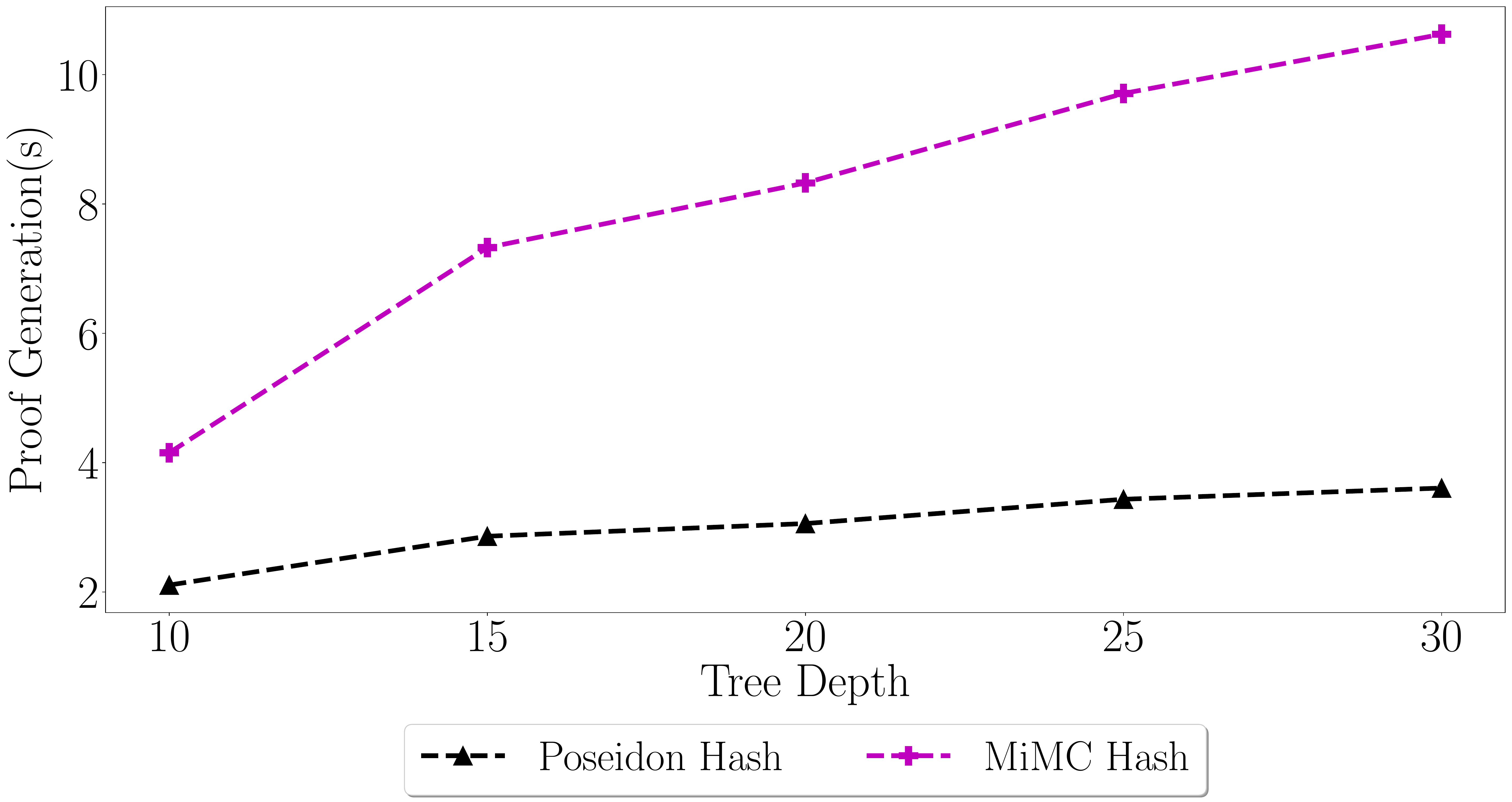}
    \caption{zkSnark Proof Generation Time for Poseidon and MiMC hash functions.
    }
    \label{fig:client-cost}
    \end{minipage}
\end{figure}

\pparagraph{Onchain Costs}
\Cref{fig:onchain-cost} provides the overall costs of deployment, deposit, reward, and withdraw for different tree depths. 
The cost of deploying the contract is the most expensive operation, 
accounting from $\approx 6m$ gas for $h=10$ to $\approx 8m$ gas for $h=30$ for both the MiMC and Poseidon hash functions.
However, we note that the deployment cost is a one-time cost which is amortized
over the lifetime of the contract.
The cost of the depositing transaction depends on the depth of the tree, which
is approximately $43,000+51,000\times h$ for the MiMC hash and approximately
$43,000+41,000\times h$ for the Poseidon hash function. The gas cost for
verifying a withdrawing transaction is approximately $320,000$ for all tree
depths and both choices of hash functions. 
The gas cost for a reward-redeeming transaction is equal to a total gas cost
of a deposit and a withdraw as the \system contract needs to verify the 
zkSnark proof as well as to update the Merkle tree.

\pparagraph{\zksnark Proof Generation} 
As the Poseidon hash function generates less constraints for the arithmetic
circuit, than the MiMC hash function (i.e.\ $243$ vs $1323$), we observe a
reduction of $3\times$ for the clients' proof generation time with an \system
system using the Poseidon hash function.
\Cref{fig:client-cost} presents the time for a client to generate the zkSnark
proofs for different tree depths and hash functions. 

\pparagraph{Lending Platforms' Additional Costs} 
In additional to the cost of executing cryptographic functions in the \system
contract, we also need to consider the cost of other interactions with
decentralized lending platforms such as Aave~\cite{aave} or
Compound~\cite{compound}. These costs are the gas cost of depositing into and
redeeming from lending platforms. 
We estimate the costs of these interactions using data from
Etherscan~\footnote{\url{https://etherscan.io/}} and Compound developer
documentations \footnote{\url{https://compound.finance/docs\#networks}}.
Thus, depositing into these lending platforms takes approximately $0.3m$ gas
(for both Aave and Compound), and redeeming from these platforms takes less
than $0.2m$ gas for Aave and less than $0.1m$ gas for Compound. 
Therefore, depending on the choice of lending platforms, 
we would expect additional $0.3m$ gas for \system's depositing function and
additional $0.2m$ gas for \system's withdrawing function.

\subsection{Empirical analysis on Tornado Cash}
\begin{figure}[t]
    \begin{minipage}{\columnwidth}
    \centering
    \includegraphics[width=\columnwidth]{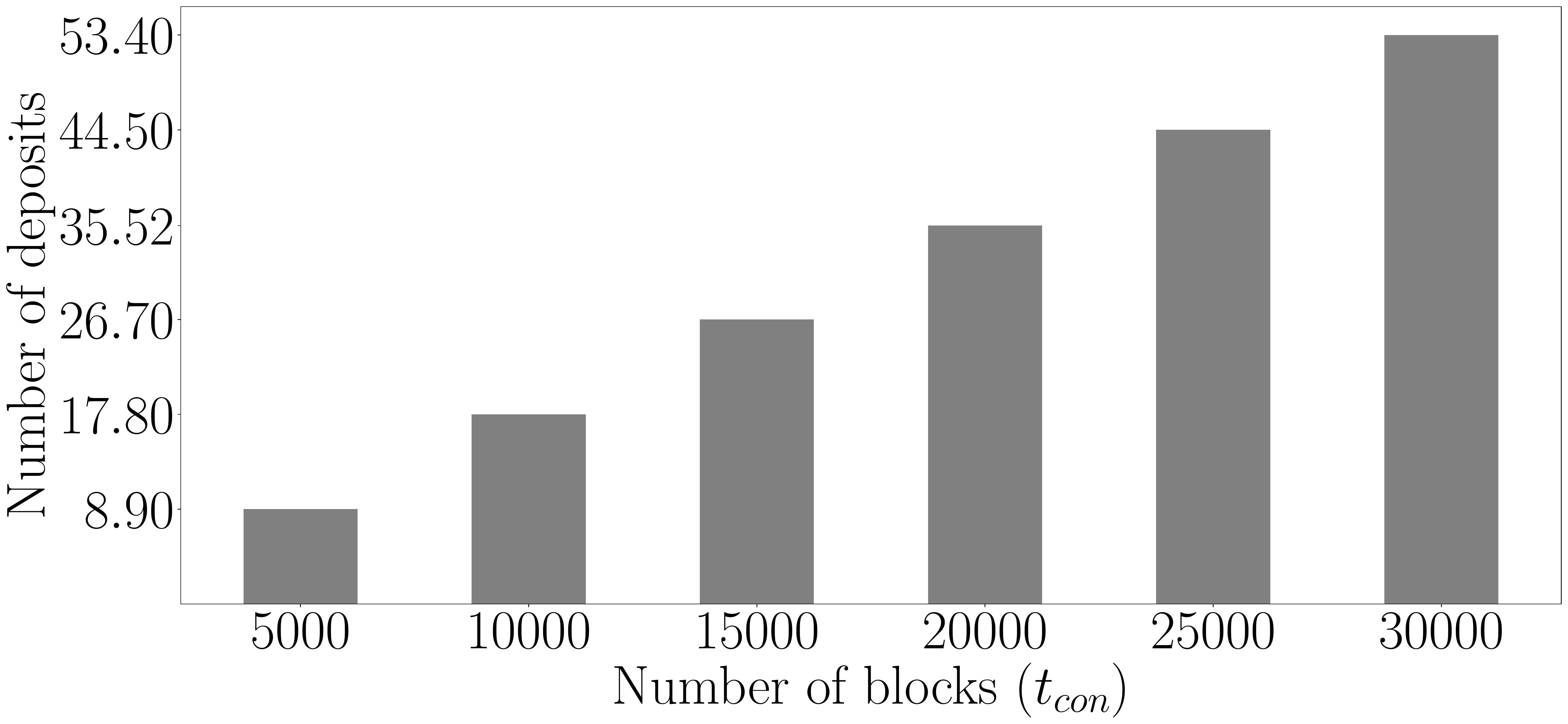}
    \caption{Average number of deposit transactions issued to the contract 
over the span of $5,000$, $10,000$, $15,000$, $20,000$, $25,000$, $30,000$ blocks.}
    \label{fig:averageNumDep}
    \end{minipage}
\end{figure}
To become eligible for a reward payment in \system, clients need to keep their deposit in the contract locked for a predefined period (i.e.\ $t_{con}$ blocks). Thus, one needs to decide what the suitable value for $t_{con}$ is (this value could be set by voting with the governance token).

We perform an empirical analysis
on the tornado cash system~\cite{Tornado} which is, to the best of our knowledge, the only \zksnark-based mixer deployed to the Ethereum main net.
Tornado cash supports two operations: deposit and withdraw.
We analyzed their $10.0$ ETH denomination deposit pool~\footnote{Address: 0x910Cbd523D972eb0a6f4cAe4618aD62622b39DbF} 
from block $9,161,895$ (25 December 2019) to block $10,726,597$ (25 August 2020) to understand how frequent clients deposit to the tornado cash system.
This frequency allows us to derive an appropriate value for how long client should keep their funds
in \system contract to be eligible for a reward.
For example, \Cref{fig:averageNumDep} suggests that for the waiting period of $t_{con}=30,000$ (approximately $4.5$ days), we can expect an additional $52$ deposit transactions issued to the contract intermittently,  
and the more deposit transactions reach the contract, the higher the anonymity set becomes. 

\begin{figure}[b]
    \begin{minipage}{\columnwidth}
    \centering
    \includegraphics[width=\columnwidth]{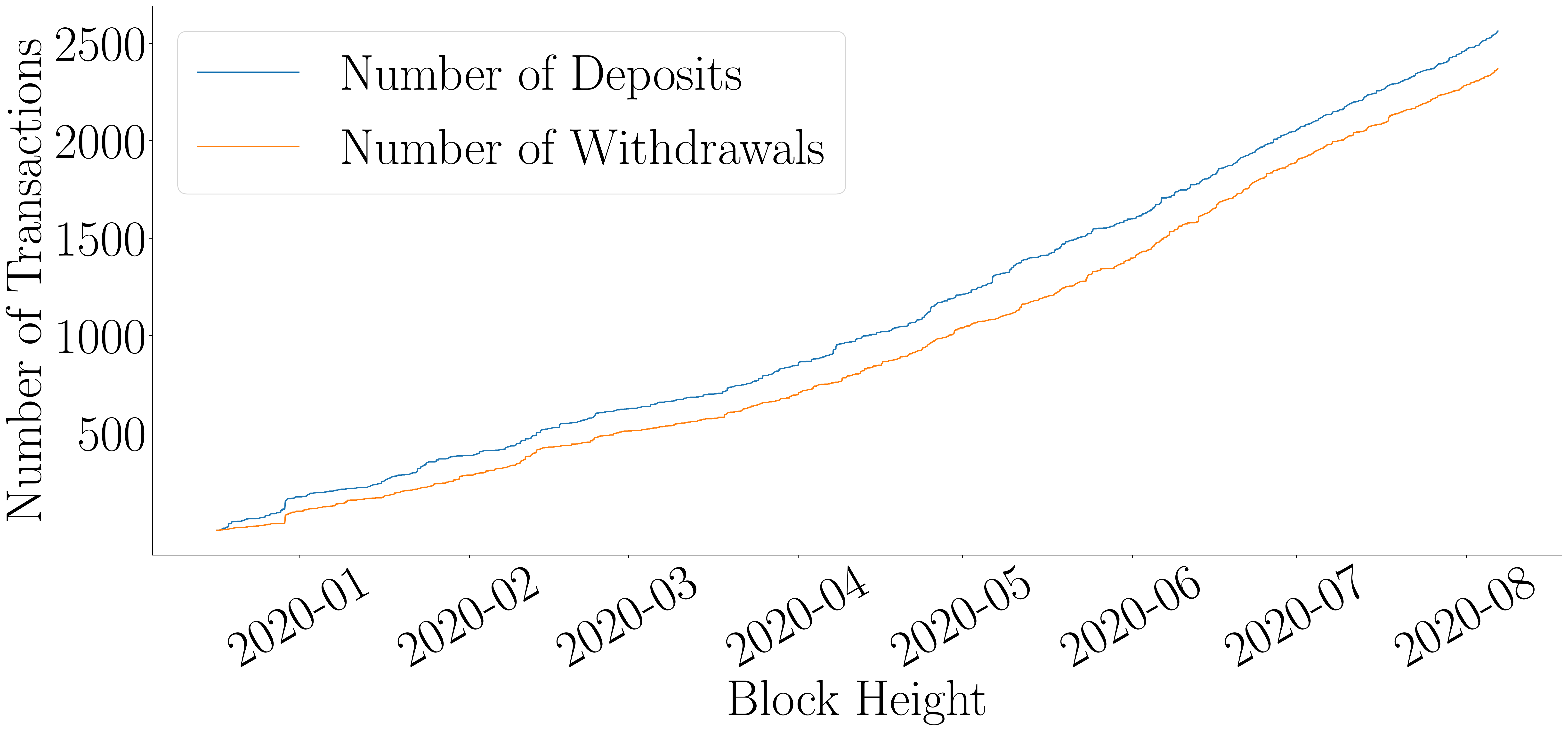}
    \caption{Number of deposits and withdrawals issued to the tornado cash $10$ ETH pool.}
    \label{fig:distanceovertime}
    \end{minipage}
\end{figure}

Moreover, over the course of $8$ months (Cf. \Cref{fig:distanceovertime}), we observe a total of $2,810$ deposit transactions, and $2,606$ withdrawing transactions on the tornado cash contract.
We note that if the number of withdrawing transactions equals to the number 
of deposit transactions at any point in time,
the size of the anonymity set is reduced to zero. 
Thus, in contrast to Tornado cash, the reward mechanism in \system is used to
incentivise clients to keep a deposit in the system to help maintain a healthy
gap between the number of deposits and withdrawals.


\section{Related Work on add-on privacy solutions}
\label{sec:relatedwork}

\pparagraph{Add-on privacy solutions for smart contract-enabled blockchains}
While we are not aware of any academic works that propose a \zksnark-based mixing system as ours, 
Tornado cash~\cite{Tornado} appears to be the first system deployed in production which allows clients to 
deposit and withdraw fixed amount of coins. 
Our work is to the best of our knowledge the first academic work which presents 
such a system, formalizes the privacy and security properties, and importantly, 
adds a novel privacy preserving reward mechanism.

Meiklejohn \emph{et al.}~\cite{meiklejohn2018mobius} propose an Ethereum-based
tumbler called M\"{o}bius. 
The construction of M\"{o}bius relies on the linkable ring signature primitive and stealth address 
mechanism used in Monero~\cite{zero-to-monero} to hide the address of the true sender and the recipient. 
However, in M\"{o}bius, the size of the anonymity set is limited to the size of the ring,
and the gas cost of the withdrawing transaction increases linearly with the size of the ring.
Thus, in term of privacy, \system offers a bigger anonymity set over time while operating 
at constant system costs.

B\"{u}nz \emph{et al.} proposed a private payment protocol for the Ethereum blockchain called Zether~\cite{zether-bunz-2020}.
The core idea of Zether is to use Elgamal to encrypt the balances of clients. 
However, the cost of Zether transactions (i.e.\ $7.8m$ gas) is expensive for Ethereum, and Zether does not hide the receiver and recipient of a transaction. 
Diamond proposed Anonymous Zether~\cite{anonymous-zether-2020} to address the later drawback, but the cost of Anonymous Zether is still expensive for blockchains such as Ethereum. For the maximum anonymity set of size $64$ reported in the paper, the gas cost of a single transferring call in Anonymous Zether is $48.7m$ gas which is approximately $32$ times the cost of an \system deposit and $130$ times the cost of an \system withdrawal for $h=30$ (The block gas limit in Ethereum is about $15m$ gas at the time of writing).

Rondelet and Zajac propose Zeth~\cite{zeth-clearmatic}, which implements all functionalities of ZeroCash~\cite{sasson2014zerocash} 
as an Ethereum smart contract. While Zeth allows expressive functionalities,
such as transferring arbitrary denomination of notes, 
it comes with the cost of using a bigger \zksnark circuit than in \system.
The choice of the SHA256 hash function in Zeth would result in approximately $59,281$ 
constraints in the arithmetic circuit, which is $50\times$ bigger than the constraints from using the MiMC hash (i.e. $1,323$ constraints) and $200\times$ bigger than the constraints from using the Poseidon hash (i.e. $243$ constraints). 
While the authors of Zeth did not report any numbers on the proof generation time, 
we expect that the \zksnark proof generation time in Zeth is an order of magnitude larger 
than in \system.
The Zeth contract needs to store all encrypted notes from 
transferring function calls, 
and depending on the size of the transaction, this additional storage
also incurs cost (storing a $32$-byte data costs $20,000$ gas~\cite{wood2014ethereum}).
The authors of Zeth also report that an estimated cost of verifying a \zksnark proof is approximately $2m$ gas ($5 \times$ the cost in \system).



\pparagraph{Other Tumbler Designs} The community proposes several centralised
tumbler
designs~\cite{bonneau2014mixcoin,valenta2015blindcoin,heilman2017tumblebit,tairi2019a2l}.
The main essence of those designs relies on a centralised offchain server to mix users' funds,
e.g., Tumblebit~\cite{heilman2017tumblebit} and A2L~\cite{tairi2019a2l}. Both
require less trust in the offchain server than solutions such as
Mixcoin~\cite{bonneau2014mixcoin} and Blindcoin~\cite{valenta2015blindcoin} by
preventing the server from stealing funds from participants. 
However, centralised
tumbling protocols cannot ensure the availability property,
because the centralised system can always censor deposits from clients.

Existing decentralized tumbler designs, such as Coinshuffle~\cite{ruffing2014coinshuffle,ruffing2017coinshufflepp}
and Coinjoin~\cite{maxwell2013coinjoin}, address the availability problem by 
proposing protocols allowing participants to interact and form transactions
that helps hide the sender from the recipient.
However, the availability of participants and the interactivity among them can 
be difficult to enforce and may lead to privacy leaking side channels.

\section{Conclusion}
\label{sec:conclusion}

Coin mixers allow alleviating to some degree the missing privacy properties of
open and permissionless blockchains. Their operations are cost-intensive both
from a transaction fee perspective and because ``better'' privacy is more
expensive than ``weaker'' privacy when measuring privacy quality quantitatively
with the anonymity set size.

In this work, we introduce a \zksnark-based coin mixer \system. \system is to
our knowledge the first construction that allows to reward mixer participants
which hold coins within the mixer for at least time $t$. Moreover, \system
allows users to earn interest on the deposited funds by leveraging popular DeFi
lending platforms. This incentive mechanism should not only attract
privacy-seeking users, but also participants that are interested in the
underlying reward distribution. Therefore, we hope that such a system
fundamentally broadens the diversity of the mixer user, improving the anonymity
set quality for all involved users. Our implementation and evaluation shows
that our mixer is practical by supporting anonymity set sizes beyond thousands
of users. 
We invite the interested reader 
to observe our short demo at \url{https://youtu.be/-oAQlsRTF08}.

{
\bibliographystyle{plain}
\bibliography{references}

\begin{thebibliography}{10}

\bibitem{aave}
Aave: The money market protocol.
\newblock \url{https://aave.com/}.

\bibitem{yield-farming}
Compound.
\newblock Available at: \url{https://compound.finance/}.

\bibitem{compound}
Compound.
\newblock \url{https://compound.finance/}.

\bibitem{curve}
Curve dao.
\newblock \url{https://curve.fi/}.

\bibitem{inefficency-sha256}
Jubjub.
\newblock Available at: \url{https://z.cash/technology/jubjub/}.

\bibitem{Tornado}
Tornado cash.
\newblock Available at: \url{https://tornado.cash/}.

\bibitem{yearn}
{Y}earn finance.
\newblock \url{https://yearn.finance/}.

\bibitem{zcash}
Zcash.
\newblock Available at: \url{https://z.cash/}.

\bibitem{MiMC-hash}
Martin Albrecht, Lorenzo Grassi, Christian Rechberger, Arnab Roy, and Tyge
  Tiessen.
\newblock Mimc: Efficient encryption and cryptographic hashing with minimal
  multiplicative complexity.
\newblock In Jung~Hee Cheon and Tsuyoshi Takagi, editors, {\em Advances in
  Cryptology -- ASIACRYPT 2016}, pages 191--219, Berlin, Heidelberg, 2016.
  Springer Berlin Heidelberg.

\bibitem{zero-to-monero}
Kurt~M. Alonso.
\newblock {Zero} to {Monero}: First edition. a technical guide to a private
  digital currency; for beginners, amateurs, and experts.
\newblock \url{https://web.getmonero.org/library/Zero-to-Monero-2-0-0.pdf}.

\bibitem{androulaki2013evaluating}
Elli Androulaki, Ghassan~O Karame, Marc Roeschlin, Tobias Scherer, and Srdjan
  Capkun.
\newblock Evaluating user privacy in bitcoin.
\newblock In {\em International Conference on Financial Cryptography and Data
  Security}, pages 34--51. Springer, 2013.

\bibitem{bano2019sok}
Shehar Bano, Alberto Sonnino, Mustafa Al-Bassam, Sarah Azouvi, Patrick McCorry,
  Sarah Meiklejohn, and George Danezis.
\newblock Sok: Consensus in the age of blockchains.
\newblock In {\em Proceedings of the 1st ACM Conference on Advances in
  Financial Technologies}, pages 183--198, 2019.

\bibitem{secure-mpc-sp2015}
E.~{Ben-Sasson}, A.~{Chiesa}, M.~{Green}, E.~{Tromer}, and M.~{Virza}.
\newblock Secure sampling of public parameters for succinct zero knowledge
  proofs.
\newblock In {\em 2015 IEEE Symposium on Security and Privacy}, pages 287--304,
  2015.

\bibitem{rsa-acc-benaloh-2007}
Josh Benaloh and Michael de~Mare.
\newblock One-way accumulators: A decentralized alternative to digital
  signatures.
\newblock In Tor Helleseth, editor, {\em Advances in Cryptology --- EUROCRYPT
  '93}, pages 274--285, Berlin, Heidelberg, 1994. Springer Berlin Heidelberg.

\bibitem{boneh-shoup-book-2020}
Dan Boneh and Victor Shoup.
\newblock A graduate course in applied cryptography, 2020.

\bibitem{bonneau2015sok}
Joseph Bonneau, Andrew Miller, Jeremy Clark, Arvind Narayanan, Joshua~A Kroll,
  and Edward~W Felten.
\newblock Sok: Research perspectives and challenges for bitcoin and
  cryptocurrencies.
\newblock In {\em Symposium on Security and Privacy}, pages 104--121. IEEE,
  2015.

\bibitem{bonneau2014mixcoin}
Joseph Bonneau, Arvind Narayanan, Andrew Miller, Jeremy Clark, Joshua~A Kroll,
  and Edward~W Felten.
\newblock Mixcoin: Anonymity for bitcoin with accountable mixes.
\newblock In {\em International Conference on Financial Cryptography and Data
  Security}, pages 486--504. Springer, 2014.

\bibitem{mpc-generating-zksnark-fc2018}
Sean Bowe, Ariel Gabizon, and Matthew~D. Green.
\newblock A multi-party protocol for constructing the public parameters of the
  pinocchio zk-snark.
\newblock In Aviv Zohar, Ittay Eyal, Vanessa Teague, Jeremy Clark, Andrea
  Bracciali, Federico Pintore, and Massimiliano Sala, editors, {\em Financial
  Cryptography and Data Security}, pages 64--77, Berlin, Heidelberg, 2019.
  Springer Berlin Heidelberg.

\bibitem{mpc-setup-bowe-miller-2017}
Sean Bowe, Ariel Gabizon, and Ian Miers.
\newblock Scalable multi-party computation for zk-snark parameters in the
  random beacon model.
\newblock Cryptology ePrint Archive, Report 2017/1050, 2017.
\newblock \url{https://eprint.iacr.org/2017/1050}.

\bibitem{zether-bunz-2020}
Benedikt B{\"{u}}nz, Shashank Agrawal, Mahdi Zamani, and Dan Boneh.
\newblock Zether: Towards privacy in a smart contract world.
\newblock {\em {IACR} Cryptol. ePrint Arch.}, 2019:191, 2019.

\bibitem{lego-snark-ccs-2019}
Matteo Campanelli, Dario Fiore, and Ana\"{\i}s Querol.
\newblock Legosnark: Modular design and composition of succinct zero-knowledge
  proofs.
\newblock In {\em Proceedings of the 2019 ACM SIGSAC Conference on Computer and
  Communications Security}, CCS '19, page 2075–2092, New York, NY, USA, 2019.
  Association for Computing Machinery.

\bibitem{marlin-eurocrypt-2020}
Alessandro Chiesa, Yuncong Hu, Mary Maller, Pratyush Mishra, Noah Vesely, and
  Nicholas Ward.
\newblock Marlin: Preprocessing zksnarks with universal and updatable srs.
\newblock In Anne Canteaut and Yuval Ishai, editors, {\em Advances in
  Cryptology -- EUROCRYPT 2020}, pages 738--768, Cham, 2020. Springer
  International Publishing.

\bibitem{anonymous-zether-2020}
Benjamin~E. Diamond.
\newblock "many-out-of-many" proofs with applications to anonymous zether.
\newblock Cryptology ePrint Archive, Report 2020/293, 2020.
\newblock \url{https://eprint.iacr.org/2020/293}.

\bibitem{eskandari-sok-frontrun-fc2019}
Shayan Eskandari, Seyedehmahsa Moosavi, and Jeremy Clark.
\newblock Sok: Transparent dishonesty: Front-running attacks on blockchain.
\newblock In Andrea Bracciali, Jeremy Clark, Federico Pintore, Peter~B.
  R{\o}nne, and Massimiliano Sala, editors, {\em Financial Cryptography and
  Data Security}, pages 170--189, Cham, 2020. Springer International
  Publishing.

\bibitem{plonk-2019}
Ariel Gabizon, Zachary~J. Williamson, and Oana Ciobotaru.
\newblock Plonk: Permutations over lagrange-bases for oecumenical
  noninteractive arguments of knowledge.
\newblock Cryptology ePrint Archive, Report 2019/953, 2019.
\newblock \url{https://eprint.iacr.org/2019/953}.

\bibitem{gervais2014privacy}
Arthur Gervais, Srdjan Capkun, Ghassan~O Karame, and Damian Gruber.
\newblock On the privacy provisions of bloom filters in lightweight bitcoin
  clients.
\newblock In {\em Computer Security Applications Conference}, pages 326--335,
  2014.

\bibitem{poseidon-hash-2019}
Lorenzo Grassi, Dmitry Khovratovich, Christian Rechberger, Arnab Roy, and
  Markus Schofnegger.
\newblock Poseidon: A new hash function for zero-knowledge proof systems.
\newblock Cryptology ePrint Archive, Report 2019/458, 2019.
\newblock \url{https://eprint.iacr.org/2019/458}.

\bibitem{groth-zksnark-2016}
Jens Groth.
\newblock On the size of pairing-based non-interactive arguments.
\newblock In Marc Fischlin and Jean-S{\'e}bastien Coron, editors, {\em Advances
  in Cryptology -- EUROCRYPT 2016}, pages 305--326, Berlin, Heidelberg, 2016.
  Springer Berlin Heidelberg.

\bibitem{heilman2017tumblebit}
Ethan Heilman, Leen Alshenibr, Foteini Baldimtsi, Alessandra Scafuro, and
  Sharon Goldberg.
\newblock Tumblebit: An untrusted bitcoin-compatible anonymous payment hub.
\newblock In {\em Network and Distributed System Security Symposium}, 2017.

\bibitem{circom-lib}
Iden3.
\newblock {C}ircom: Circuit compiler for zksnark.
\newblock \url{https://github.com/iden3/snarkjs}.

\bibitem{snarkjs-lib}
Iden3.
\newblock {S}narkjs: Javascript and pure web assembly implementation of zksnark
  schemes.
\newblock \url{https://github.com/iden3/snarkjs}.

\bibitem{rsa-acc-li-2007}
Jiangtao Li, Ninghui Li, and Rui Xue.
\newblock Universal accumulators with efficient nonmembership proofs.
\newblock In Jonathan Katz and Moti Yung, editors, {\em Applied Cryptography
  and Network Security}, pages 253--269, Berlin, Heidelberg, 2007. Springer
  Berlin Heidelberg.

\bibitem{sonic-ccs-2019}
Mary Maller, Sean Bowe, Markulf Kohlweiss, and Sarah Meiklejohn.
\newblock Sonic: Zero-knowledge snarks from linear-size universal and updatable
  structured reference strings.
\newblock In {\em Proceedings of the 2019 ACM SIGSAC Conference on Computer and
  Communications Security}, CCS '19, page 2111–2128, New York, NY, USA, 2019.
  Association for Computing Machinery.

\bibitem{maxwell2013coinjoin}
Greg Maxwell.
\newblock Coinjoin: Bitcoin privacy for the real world.
\newblock In {\em Post on Bitcoin forum}, 2013.

\bibitem{meiklejohn2018mobius}
Sarah Meiklejohn and Rebekah Mercer.
\newblock M{\"o}bius: Trustless tumbling for transaction privacy.
\newblock {\em Proceedings on Privacy Enhancing Technologies},
  2018(2):105--121, 2018.

\bibitem{Merkle1988}
Ralph~C Merkle.
\newblock A digital signature based on a conventional encryption function.
\newblock In {\em Conference on the theory and application of cryptographic
  techniques}, pages 369--378. Springer, 1987.

\bibitem{pedersen-hash}
Silvio Micali, Michael Rabin, and Joe Kilian.
\newblock Zero-knowledge sets.
\newblock In {\em Proceedings of the 44th Annual IEEE Symposium on Foundations
  of Computer Science}, FOCS ’03, page~80, USA, 2003. IEEE Computer Society.

\bibitem{miers2013zerocoin}
Ian Miers, Christina Garman, Matthew Green, and Aviel~D Rubin.
\newblock Zerocoin: Anonymous distributed e-cash from bitcoin.
\newblock In {\em Symposium on Security and Privacy}, pages 397--411, 2013.

\bibitem{zcash-2017-mpc-setup}
Andrew Miller and Sean Bowe.
\newblock {Zcash MPC Setup}.
\newblock \url{https://www.zfnd.org/blog/powers-of-tau/}.

\bibitem{Rogaway2004}
Phillip Rogaway and Thomas Shrimpton.
\newblock Cryptographic hash-function basics: Definitions, implications, and
  separations for preimage resistance, second-preimage resistance, and
  collision resistance.
\newblock In {\em FSE 2004}, pages 371--388, 2004.

\bibitem{zeth-clearmatic}
Antoine Rondelet and Michal Zajac.
\newblock Zeth: On integrating zerocash on ethereum, 2019.

\bibitem{ruffing2017valueshuffle}
Tim Ruffing and Pedro Moreno-Sanchez.
\newblock Valueshuffle: Mixing confidential transactions for comprehensive
  transaction privacy in bitcoin.
\newblock In {\em International Conference on Financial Cryptography and Data
  Security}, pages 133--154. Springer, 2017.

\bibitem{ruffing2014coinshuffle}
Tim Ruffing, Pedro Moreno-Sanchez, and Aniket Kate.
\newblock Coinshuffle: Practical decentralized coin mixing for bitcoin.
\newblock In {\em European Symposium on Research in Computer Security}, pages
  345--364. Springer, 2014.

\bibitem{ruffing2017coinshufflepp}
Tim Ruffing, Pedro Moreno{-}Sanchez, and Aniket Kate.
\newblock {P2P} mixing and unlinkable bitcoin transactions.
\newblock In {\em Network and Distributed System Security Symposium}, 2017.

\bibitem{sasson2014zerocash}
Eli~Ben Sasson, Alessandro Chiesa, Christina Garman, Matthew Green, Ian Miers,
  Eran Tromer, and Madars Virza.
\newblock Zerocash: Decentralized anonymous payments from bitcoin.
\newblock In {\em Symposium on Security and Privacy}, pages 459--474. IEEE,
  2014.

\bibitem{tairi2019a2l}
Erkan Tairi, Pedro Moreno-Sanchez, and Matteo Maffei.
\newblock A2l: Anonymous atomic locks for scalability and interoperability in
  payment channel hubs.
\newblock Technical report, Cryptology ePrint Archive, Report 2019/589, 2019.

\bibitem{valenta2015blindcoin}
Luke Valenta and Brendan Rowan.
\newblock Blindcoin: Blinded, accountable mixes for bitcoin.
\newblock In {\em International Conference on Financial Cryptography and Data
  Security}, pages 112--126. Springer, 2015.

\bibitem{wood2014ethereum}
Gavin Wood.
\newblock Ethereum: A secure decentralised generalised transaction ledger.
\newblock {\em Ethereum project yellow paper}, 151:1--32, 2014.

\end{thebibliography}
}

\appendix

\section{Discussion and Applications}
\label{sec:discussion}
\pparagraph{Trusted Setup in \zksnark}
As discussed in~\Cref{sec:preliminaries}, a \zksnark requires a trusted setup 
to generate the evaluation and proving key for each circuit.
While one can assume that there exists a trusted third party which helps run
the setup, this trust assumption is typically not welcome by the blockchain
community, because if such third party can maliciously generate the keys (or
the common reference string), it can form a valid proof and steal the contract
funds.

To remove the trusted third party assumption, one can run a multi-party computation
(MPC) setup where users can contribute a share to the trusted setup.
Several works~\cite{mpc-generating-zksnark-fc2018,secure-mpc-sp2015,mpc-setup-bowe-miller-2017} proposed different protocols for such trusted setup, 
and they showed that as long as one participant is honest, the \zksnark instance will be secure. 
In particular, the Zcash team has performed such MPC setup for their protocol parameters in 2017~\cite{zcash-2017-mpc-setup}. 
However, the MPC setup may need to be carried out independently for different circuits and related works~\cite{sonic-ccs-2019,lego-snark-ccs-2019,plonk-2019,marlin-eurocrypt-2020} have proposed 
several \zksnark constructions that utilizes a universal setup that can be used for \emph{any} circuits with bounded size. 
These \zksnark constructions can be easily integrated into \system in the future.

\pparagraph{Transferring arbitrary denomination} 
The current version of \system does not allow clients to transfer arbitrary amount of coins
privately among clients. 
To achieve such property, one either needs an out-of-band communication channel between
a sender and a recipient to transfer secret notes, or the sender can spend more fees to 
store additional encrypted data onchain. 
Moreover, to prevent a sender from stealing coins from the recipient, 
one could use a similar commitment scheme and encryption as used in Zcash~\cite{zcash};
however, the use of these primitives will increase the cost of the onchain verification. 
Nevertheless, we will leave the transferring functionality of \system for the future work. 

\pparagraph{Sender outsources transaction fee payment}
Issuing a transaction requires the payment of fees, and clients should not use
the same address for such payment, otherwise their addresses can be linked. In
practice users can chose to use a relayer, who broadasts transactions and is
paid from a fraction of the withdraw or reward transaction. The relayer can
receive the corresponding client proof through a side channel.

\pparagraph{Constant querying state} Most blockchain clients (e.g.\ MetaMask) outsource their blockchain information to centralized services such as Infura. Those centralized services are aware of, the clients' blockchain address(es), IP address as well of the fact that the client queried the \system contract state. These services are therefore privacy critical, as they may be able to link different addresses from the same client. We hence recommend a privacy aware client to operate an independent validating full blockchain client, or use network-level anonymity solutions such as Tor or Virtual Private Network (VPN) before connecting to these centralized services.

\pparagraph{Decentralized governance}
Once deployed, \system's system parameters, will likely need to be adjusted
during its lifetime. One could chose an admin key to govern \system, for the
sake of decentralization, however, we believe that a decentralized approach
would be beneficial.
A governance token is hence the natural choice, whereby \system can itself
distribute those tokens to the clients participating in the protocol. We have
identified the following parameters that should be governed: \emph{(i)} new
relayer addresses, \emph{(ii)} condition for client reward, \emph{(iii)} the
amount of the reward. Once a new version of \system is developed, the
governance mechanism could vote to \emph{(iv)} migrate deposits to a new
contract with new features/bug fixes.

\end{document}